\documentclass[preprint,12pt]{elsarticle}
\usepackage{subcaption}
\usepackage{hyperref}
\usepackage{algorithm,algorithmic}
\usepackage{booktabs, multirow}
\usepackage{rotating}
\usepackage{enumitem}

\usepackage{amssymb}
\usepackage{amsmath}

\usepackage{tipa} 

\usepackage{cancel}
\usepackage{soul} 
\usepackage{xcolor} 

\usepackage[final]{changes}

\journal{Computer Speech and Language}

\begin{document}

\begin{frontmatter}



\title{Multimodal Laryngoscopic Video Analysis for Assisted Diagnosis of Vocal Fold Paralysis}


\author[1,2]{Yucong Zhang$^\ast$} 
\author[3,5]{Xin Zou$^\ast$\begingroup
\renewcommand{\thefootnote}{}
\footnotetext{$\ast$ The first two authors contribute equally to this work.}
\endgroup}
\author[3]{Jinshan Yang}
\author[3]{Wenjun Chen}
\author[1,4]{Juan Liu}
\author[3]{Faya Liang$^\dagger$}
\author[1,2]{Ming li$^\dagger$\begingroup
\renewcommand{\thefootnote}{}
\footnotetext{$\dagger$ Corresponding authors: Faya Liang (liangfy3@mail.sysu.edu.cn) and Ming Li (ming.li369@dukekunshan.edu.cn)}
\endgroup}

\affiliation[1]{organization={School of Computer Science},
            addressline={Wuhan University}, 
            city={Wuhan},
            postcode={430072}, 
            state={HuBei},
            country={China}}
\affiliation[2]{organization={Suzhou Municipal Key Laboratory of  Multimodal Intelligent Systems},
            addressline={Duke Kunshan University}, 
            city={Suzhou},
            postcode={215316}, 
            state={Jiangsu},
            country={China}}
\affiliation[3]{organization={Sun Yat-sen Memorial Hospital of Sun Yat-sen University},
            city={Guangzhou},
            postcode={510000}, 
            state={Guangdong},
            country={China}}
\affiliation[4]{organization={School of Artificial Intelligent},
            addressline={Wuhan University}, 
            city={Wuhan},
            postcode={430072}, 
            state={HuBei},
            country={China}}
\affiliation[5]{organization={Department of Otorhinolaryngology},
            addressline={Shenzhen Children’s Hospital}, 
            postcode={518000}, 
            state={Shenzhen},
            country={China}}

\begin{abstract}
This paper presents the Multimodal Laryngoscopic Video Analyzing System~(MLVAS)\footnote{Demo available at \url{https://drive.google.com/file/d/15on-o7BGB91jwOnz3P2BqzXEhVsTkBjz/view?usp=sharing}}, a novel system that leverages both audio and video data to automatically extract key video segments and metrics from raw laryngeal videostroboscopic videos for assisted clinical assessment. The system integrates video-based glottis detection with an audio keyword spotting method to analyze both video and audio data, identifying patient vocalizations and refining video highlights to ensure optimal inspection of vocal fold movements. Beyond key video segment extraction from the raw laryngeal videos, MLVAS is able to generate effective audio and visual features for Vocal Fold Paralysis~(VFP) detection. 
Pre-trained audio encoders are utilized to encode the patient voice to get the audio features. Visual features are generated by measuring the angle deviation of both the left and right vocal folds to the estimated glottal midline on the segmented glottis masks. To get better masks, we introduce a diffusion-based refinement that follows traditional U-Net segmentation to reduce false positives. We conducted several ablation studies to demonstrate the effectiveness of each module and modalities in the proposed MLVAS. The experimental results on a public segmentation dataset show the effectiveness of our proposed segmentation module. In addition, unilateral VFP classification results on a real-world clinic dataset demonstrate MLVAS’s ability of providing reliable and objective metrics as well as visualization for assisted clinical diagnosis.
\end{abstract}

\begin{keyword}
Glottis Segmentation, Keyword Spotting, Laryngoscope, Vocal Fold Paralysis, Pre-trained
\end{keyword}

\end{frontmatter}

\section{Introduction}
\label{sec:introduction}
Vocal fold Paralysis~(VFP) is a condition where one of the vocal folds fails to move properly, leading to voice changes, difficulty swallowing, and potential breathing problems~\cite{Wilson2021MedializationLA, Havas1999UnilateralVF}. VFP can result from nerve damage due to surgery, injury, infection, or tumors, significantly impacting a patient's quality of life~\cite{Rosenthal2007VocalFI}. Accurate diagnosis of VFP is crucial, as it informs the appropriate medical or surgical intervention, which can restore vocal function, improve airway protection, and enhance overall patient outcomes. Clinicians often use laryngeal videostroboscopy to check the vocal fold vibration in details. Laryngeal videostroboscopy is a specialized diagnostic tool used to evaluate the function of the vocal folds and the larynx. The stroboscopic component is important because it allows for visualization of vocal fold vibrations, which are often too rapid to be observed directly by doctors~\cite{casiano1992efficacy}. Stroboscopy uses a light source that flashes at a slightly different frequency than the vocal fold vibration. This creates a slow-motion effect, enabling the clinician to see the individual phases of vocal fold vibration in great details.

With the advent of the artificial intelligence, deep learning methods are developed to extract useful parameters~\cite{tao2007extracting} to assist clinicians track the motion of vocal folds~\cite{donhauser2024neural, pedersen2023localization, hamad2020ensemble, hamad2019automated, yan2006biomedical}, and even make predictions~\cite{lin2018quantification, devore2023predictive} or classifications~\cite{schwarz2006classification} to help clinicians perform diagnosis. However, the diagnosis of vocal folds requires to inspect the complete phonation cycles of patients. Previous works default to using video frames or images that already contain the vocal folds, but in most of the time, the raw recordings from the endoscopic inspection often carry useless information other than patient's phonation cycles. For instance, no phonation cycles occur in the beginning of the inspection, as the laryngoscope is still finding the vocal folds. As a result, manual labeling and selection of video frames from raw endoscopic videos are required, making the diagnosis process time-consuming and inconvenient for clinicians.

Although experienced experts can provide valuable insights through the analysis of endoscopic videos, this approach often relies on subjective diagnosis, which can compromise objectivity and increase patient anxiety, as well as heighten the risk of misdiagnosis. To address these challenges, various analytical methods have been developed to enhance the efficiency of diagnoses derived from endoscopic videos. However, most of the existing approaches rely solely on single modality, audio or video modality, neglecting the potential benefits of combining those two modalities, which could enhance diagnostic precision. 

Audio-based methodologies have exhibited remarkable potential in detecting voice pathologies. These techniques often involve transforming audio signals into sophisticated features, such as Mel spectrograms, before further modeling. For instance, Compton et al. proposed a lightweight MLP model for predicting vocal fold pathologies, claiming its performance surpassed expert evaluations~\cite{compton2023developing}. Similarly, Deep Neural Networks (DNNs) have been widely adopted in voice pathology prediction tasks~\cite{ hu2021deep, wu2018convolutional, ChuangYCHXWLF18,kim2024classification}. However, clinical data is often challenging to collect, and datasets are typically restricted from public disclosure due to patient privacy concerns, leading to a scarcity of training samples. Deep learning models, which are prone to overfitting on limited datasets, have driven researchers to use explainable handcrafted features such as Mel-frequency cepstral coefficients (MFCCs) as input for traditional machine learning methods, including SVM, RF, and NB~\cite{low2024identifying,cesarini2024multi,kim2024classification}. 
	
Another prominent solution for addressing audio data scarcity in specific tasks is the use of pre-trained audio encoders. By fine-tuning these models on small, task-specific datasets or directly use the output feature embeddings of these models, researchers have achieved comparable or even superior performance relative to state-of-the-art~(SOTA) methods on downstream applications with limited data resources. For instance, pre-trained models trained on large-scale, open-domain datasets have demonstrated SOTA performance in machine anomaly detection~\cite{jiang2024anopatch,zheng2024improving}. In healthcare, pre-training techniques have been applied to detect conditions such as Alzheimer’s disease~\cite{zhu2022domain,wang22k_interspeech} and heart failure~\cite{priyasad22_interspeech}. Specifically for voice disorder analysis, speech pre-trained models have emerged as popular pre-trained audio feature extractors. Tirronen et al. explore the effectiveness of using wav2vec~\cite{baevski2020wav2vec} feature extractor to detect voice disorders~\cite{tirronen2023utilizing,tirronen2023hierarchical}. Javanmardi et al. specifically targets the detection of the dysarthria speech using wav2vec models~\cite{javanmardi2024pre,javanmardi2024exploring}. Moreover, models pre-trained on image domain is also helpful for audio classification of voice pathologies~\cite{kim2025deep}. Despite these advancements, audio-only approaches fail to incorporate vital visual information, which provides richer and more interpretable insights into vocal fold conditions, rendering them incapable of distinguishing left-sided from right-sided paralysis. Furthermore, the scarcity of clinical audio data poses a significant challenge for developing robust and generalizable machine learning models.

Visual modalities provide intuitive and interpretable features, enabling more effective clinical diagnoses. Foundational research established a strong correlation between the maximum separation of vocal folds and VFP~\cite{inagi1997correlation}, motivating the development of methods to quantitatively analyze this separation. For instance, Gloger et al. concentrated on segmenting the glottic region by extracting physical attributes such as shape and color~\cite{gloger2014fully}. Adamian et al. and Wang et al. utilized supervised learning techniques to track vocal fold movements and calculate the Anterior Glottic Angle (AGA), which measures the angular relationship between the two vocal folds~\cite{adamian2021open,wang2021application}. These studies introduced the Glottal Area Waveform (GAW) and the AGA waveform as diagnostic tools by computing the glottal area and AGA for each video frame. Pennington et al. expanded this line of research by incorporating additional features; however, their analysis predominantly relied on metrics derived from the AGA~\cite{pennington2024development}. Despite these advances, current approaches face two key limitations: they primarily measure the separation between the left and right vocal folds, rendering them inadequate for distinguishing left-sided from right-sided paralysis, and they are restricted to pre-processed or brief video segments, often omitting sequences where vocal folds are not visible in raw endoscopic videos. These constraints hinder their application to raw, extended laryngoscopic videos commonly encountered in clinical practice, while also requiring manual annotation, which increases the workload for clinicians. 

In this article, we introduce the Multimodal Laryngoscopic Video Analyzing System (MLVAS), a novel framework that leverages both audio and visual modalities to automatically process and analyze raw laryngeal videostroboscopic recordings. The proposed system comprises a multimodal front-end module, a feature extraction module, and a multimodal back-end classification module for detecting VFP. The front-end module identifies and extracts video segments that captures complete phonation cycles with clear vocal folds' movement from raw endoscopic laryngeal videos. Subsequently, MLVAS extracts both audio and video features from these segments. By leveraging these complementary modalities, the system employs a classifier in the back-end for VFP detection. The primary contributions of this work are as follows:

\begin{itemize}
    \item[1.] We propose a multimodal system, MLVAS, that incorporates an audio-based Keyword Spotting (KWS) mechanism to automatically identify and extract relevant video segments containing complete phonation cycles and strobing sequences from raw laryngeal videostroboscopic recordings. This approach reduces the time required for clinicians to review lengthy video data.
    \item[2.] To address the challenge posed by the scarcity of clinical data, the system leverages a pre-trained audio encoder for detecting vocal fold pathologies. To the best of our knowledge, this is the first application of a pre-trained audio model for a vocal fold paralysis prediction.
    \item[3.] The proposed system further enhances VFP detection by incorporating visual features extracted from laryngeal video recordings with the audio features. Experimental results demonstrate that combining audio and video modalities leads to substantial improvements in VFP detection performance.
    \item[4.] The system introduces Left and Right Vocal Fold Dynamics (LVFDyn and RVFDyn) to reflect the vocal fold's activity, enabling precise diagnosis of left- or right-sided VFP. Experimental findings highlight its effectiveness in diagnosing Unilateral Vocal Fold Paralysis (UVFP).
\end{itemize}

The organization of this paper is as follows: Section~\ref{sec:system} introduces the proposed method for extracting key video segments, including audio keyword spotting, visual glottis detection, and strobing segment identification. Section~\ref{sec:audio_pretrain} describes the methodology for audio feature extraction using a pre-trained audio encoder. Section~\ref{sec:glottis_segmentation} outlines the two-stage glottis image segmentation technique for video feature extraction, consisting of an initial U-Net-based segmentation followed by diffusion-based refinement. Additionally, this section introduces the computation of LVFDyn and RVFDyn features derived from the segmented glottis masks. Section~\ref{sec:paralysis_analysis} details the multimodal model for detecting both VFP and UVFP. Finally, Section~\ref{sec:exp_settings} and Section~\ref{sec:exp_results} presents the experimental settings and results, and Section~\ref{sec:conclusion} concludes the paper.

\begin{figure}[t]
	\centering
	\includegraphics[width=.8\textwidth]{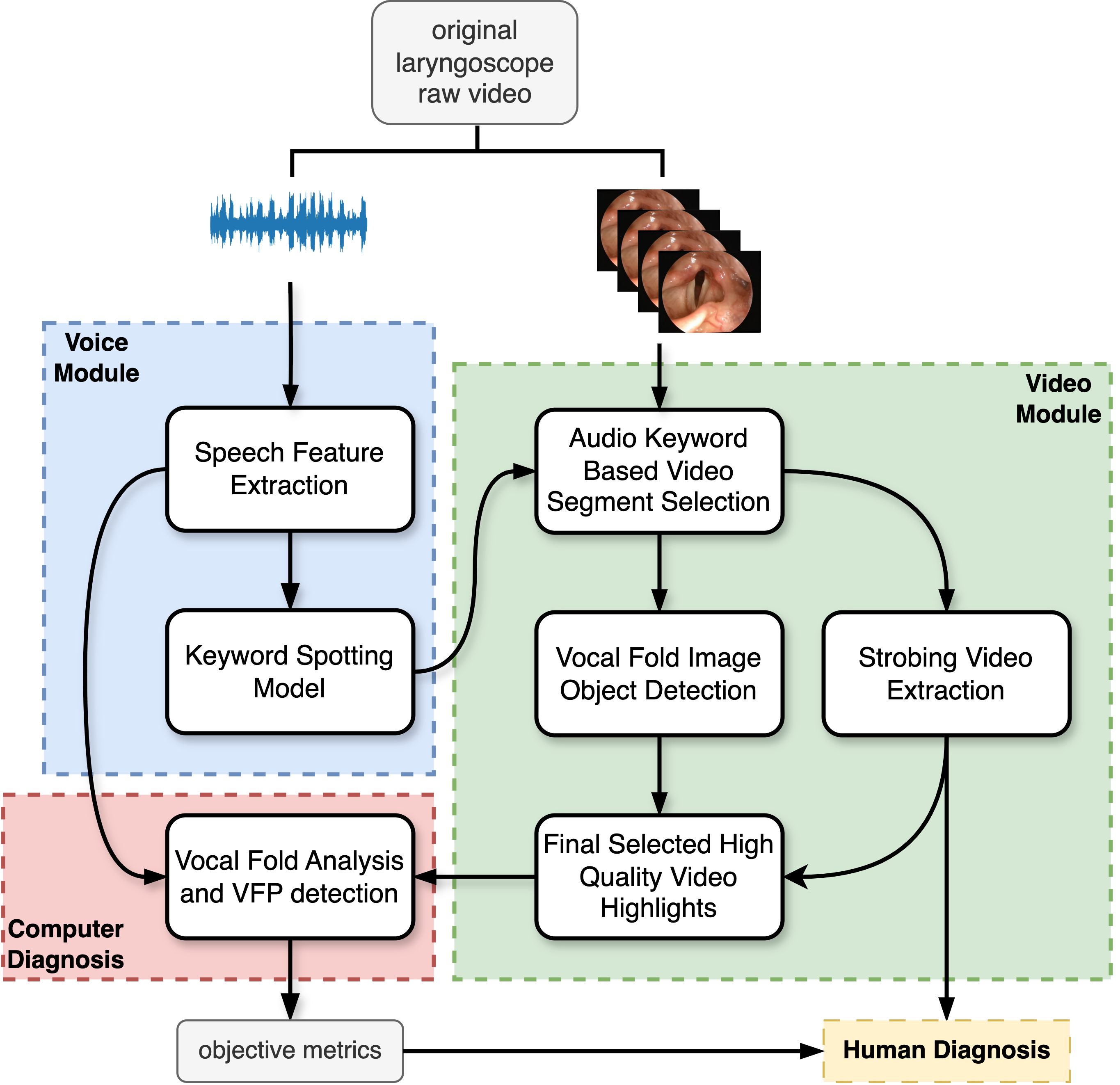}
	\caption{The overview of our proposed MLVAS framework.}
	\label{fig:system_overview}
\end{figure}

\section{Multimodal Extraction of Key Video Segments}
\label{sec:system}

\subsection{System Design}
Our system aims to facilitate efficient clinical examinations by extracting key segments from laryngoscopic videos and providing objective indicators for specific laryngeal diseases. As Figure~\ref{fig:system_overview} shows, comprising two main modules – the voice module, and the video module – our system ensures the accurate observation of vocalization cycles and the clear visualization of the glottal area.

 The voice module initially processes the audio extracted from the video using STFT to obtain spectrograms. Through the KWS technique, each frame is analyzed to detect patient vocalizations. This enables the preliminary segmentation of vocalization segments within the video. 

Subsequently, the video module further refines these vocalization segments to obtain key video segments to form laryngoscopic highlights, ensuring the visibility of the vocal folds. Specifically, by utilizing the glottis object detection model, MLVAS can identify regions containing both the vocal folds and the glottis area in each frame. Moreover, given the importance of the stroboscopic portions in laryngoscopic videos for subjective analysis by physicians, we also include a stroboscopic video extraction method into our system. 


\begin{figure*}[t]
	\centering
	\includegraphics[width=\textwidth]{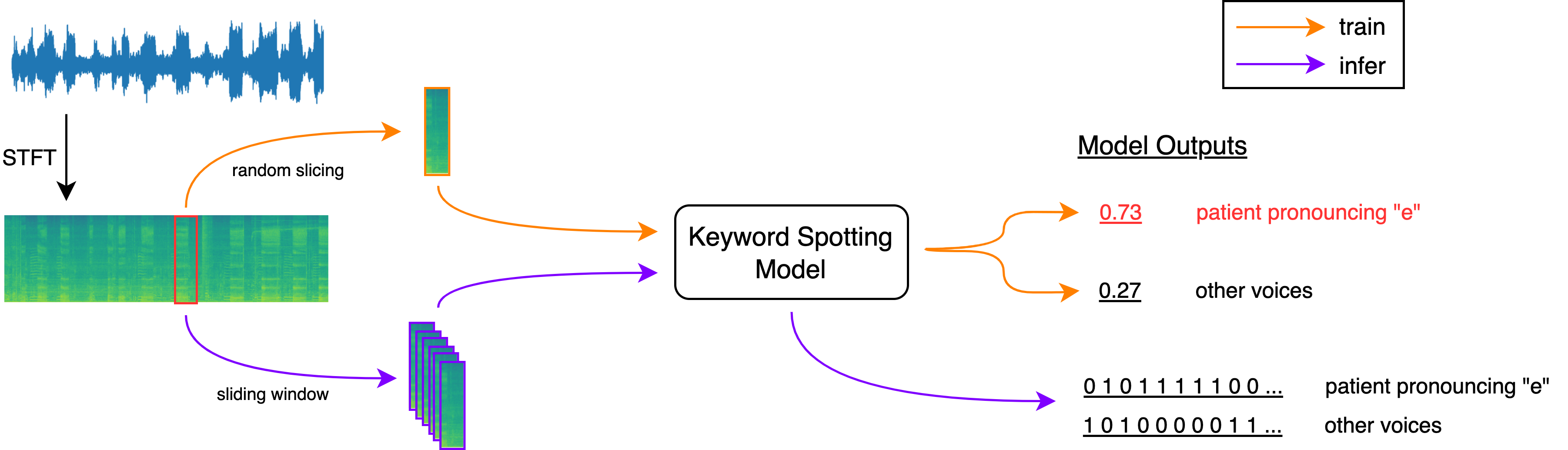}
	\caption{The overview of the audio processing module. The orange line shows the training process, and the purple line shows the inferring process.}\vspace{-4mm}
	\label{fig:kws}
\end{figure*}

\subsection{Audio Processing Module}
To get a better examination of the status of the vocal folds, \deleted{patients are often pronouncing "e" (although asked to pronounce "a") while the clinicians examine the larynx using the endoscope. In this way, the vibration of the vocal folds can be easily detected by the clinicians}\added{patients are often instructed to pronounce ``/i:/''. However, due to the physical constraints of the tongue during endoscopic procedures, the actual sound produced often resembles ``/\textepsilon:/'', which corresponds to the International Phonetic Alphabet (IPA) symbol. This sound pattern enables clinicians to detect the vibration of the vocal folds more effectively}. Hence, by capturing the audio segments that pronounce \deleted{"e"}\added{``/\textepsilon:/''}, phonation cycles can be detected \added{and analyzed}. To accomplish this, we developed a KWS model that is used to detect audio keywords in a sentence, \deleted{such as}\added{similar to those used in systems like } \deleted{"Hey Siri", "Ok Google"}\added{``Hey Siri'', ``Ok Google''}\deleted{, and in our case, the word "e".}\added{. In our case, the KWS model is designed to identify instances of the keyword ``/\textepsilon:/''.}

The overall pipeline is illustrated in Figure~\ref{fig:kws}. Initially, the input audio is transformed into a STFT spectrogram, converting the time-domain signal into a time-frequency representation for more effective analysis. The spectrogram is then segmented into chunks along the time axis, with each chunk containing a fixed number of frames. During the training phase, these chunks are randomly selected from each audio clip to ensure a diverse set of training samples. This method enhances the model's generalization capabilities by incorporating spectrograms from various audio clips into a robust batch of training data.

\begin{table}[t]
\caption{Model structure for keyword spotting}
\label{tab:keyword_spotting}
\centering
\resizebox{.7\columnwidth}{!}{\normalsize%
\begin{tabular}{|c|c|c|c|}
\hline
Operator          & \# of out channels & Kernel & stride \\ \hline
Conv2D            & 32                 & (3,3)  & (1,1)  \\ \hline
MaxPool2D         & -                  & (1,1)  & (1,1)  \\ \hline
ResNet2D          & 64                 & (3,3)  & (1,1)  \\ \hline
MaxPool2D         & -                  & (2,2)  & (2,2)  \\ \hline
ResNet2D          & 64                 & (3,3)  & (1,1)  \\ \hline
MaxPool2D         & -                  & (1,1)  & (1,1)  \\ \hline
ResNet2D          & 128                & (3,3)  & (1,1)  \\ \hline
MaxPool2D         & -                  & (2,2)  & (2,2)  \\ \hline
ResNet2D          & 128                & (3,3)  & (1,1)  \\ \hline
MaxPool2D         & -                  & (1,1)  & (1,1)  \\ \hline
AdaptiveAvgPool2D & -                  & -      & -      \\ \hline
Linear            & 32                 & -      & -      \\ \hline
Linear            & 2                  & -      & -      \\ \hline
\end{tabular}%
}
\end{table}

The spectrogram chunks are subsequently fed into the KWS model, the architecture of which is detailed in Table~\ref{tab:keyword_spotting}. The model comprises multiple convolutional blocks and residual blocks~\cite{he2016deep}, designed to efficiently extract relevant features from the spectrogram chunks. Within the model, we use max pooling to progressively reduce the spatial dimensions. Furthermore, we use an adaptive average pooling layer to aggregate the features before they pass through two fully connected layers, with the final layer producing the classification score output.

During inference, the input spectrogram is sliced into chunks using a sliding window mechanism. Each chunk is processed by the trained KWS model to generate decision results, as depicted in Figure~\ref{fig:kws}. This KWS model can reliably detects the \deleted{"e"}\added{``/\textepsilon:/''} vocalization, providing critical prior knowledge for the subsequent analysis.

\subsection{Video Processing Module}
\label{subsec:video_processing_module}

\begin{figure}[t]
	\centering
	\includegraphics[width=\textwidth]{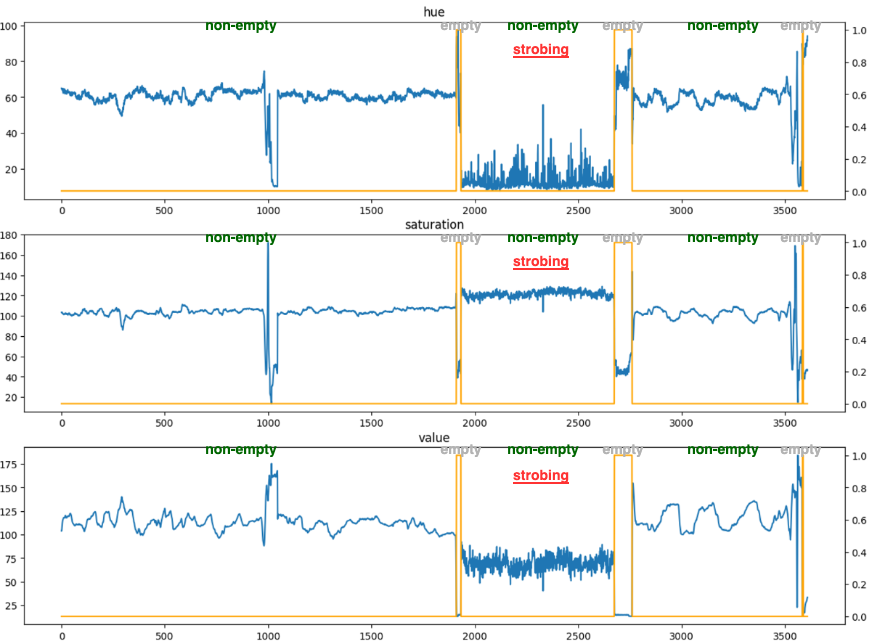}
	\caption{An example result of the HSV analysis for strobing video extraction. The blue line represents the HSV values along the time axis. The yellow line is the unit step function highlighting the empty frames with zero values.}
	\label{fig:hsv}
\end{figure}

Although the audio KWS model is employed to detect the vocalization video segment of the phonation \deleted{"e"}\added{``/\textepsilon:/''}, these detected video segments may not always successfully capture the vocal folds and glottis on the visual side. To address this issue, we train a vocal fold detection model, utilizing the famous object detection model structure YOLO-v5~\cite{yolov5} on the BAGLS dataset~\cite{gomez2020bagls}, to further refine the time masks.



As depicted in Figure~\ref{fig:system_overview}, our pipeline integrates a strobing video extraction module alongside the vocal fold image object detection module. This module is utilized to select the strobing segments of laryngeal videos by analyzing  hue, saturation, and value~(HSV) of video frames. The HSV parameters represent type, intensity, and brightness of colors, respectively. Transitions marked by empty frames typically occur at the beginning and end of strobing segments. We employ a unit step function to identify and mark all empty frames with a zero value in color~(illustrated by the yellow line in Figure~\ref{fig:hsv}). This technique segments the video into several continuous non-empty frame sequences. Within these non-empty segments, we examine the variations in HSV values. The variations is calculated as the total number of changes within a time range. Due to the rapid color-changing characteristic of strobing videos (as shown in Figure~\ref{fig:hsv}), we are able to identify the strobing segments by calculating the frequency of HSV fluctuations within each non-empty segment, selecting the segment with the highest number of fluctuations. Suppose $v_{t-1}, v_{t}, v_{t+1}$ stands for the value at time step $t-1, t$ and $t+1$ respectively. Then, the frequency of HSV fluctuation within a time range of $T=[t_0, t_n]$ is calculated as follows.
$$F_T=\sum_{t=1}^{n-1}\frac{(v_{t}-v_{t-1})\cdot(v_{t+1}-v_t)}{|(v_{t}-v_{t-1})\cdot(v_{t+1}-v_t)|}$$

\section{Audio Modeling with Pretrained Models}
\label{sec:audio_pretrain}
The core idea of our method is to utilize large-scale pre-trained models as initialization and fine-tune them on clinical data for classification tasks. In our proposed system, we employ Dasheng~\cite{dinkel24b_interspeech}, a state-of-the-art audio encoder, as the pre-trained model. Dasheng is trained on four extensive datasets encompassing speech, audio events, and music in a self-supervised manner. We selected Dasheng due to its large-scale dataset coverage, making it the most comprehensive open-source audio pre-trained model, to the best of our knowledge. Dasheng builds upon the principles of the Masked Audio Encoder (MAE)~\cite{he2022masked}, which utilizes a transformer-based asymmetric encoder-decoder architecture. Unlike the traditional audio MAE~\cite{huang2022masked}, which masks and reconstructs time-frequency patches on the Mel-spectrogram, Dasheng applies masking along the time axis in a chunk-wise manner. The model incorporates learnable absolute positional embeddings, generates frame-level embeddings at a higher frequency, and processes consecutive chunks of Mel-spectrogram frames, enabling robust and efficient feature extraction for various kinds of downstream tasks.

During fine-tuning, we initialize the model with the pre-trained weights of Dasheng. As the encoded output of Dasheng is frame-wise, we compute the mean across all frame embeddings to derive an utterance-level embedding. A linear layer is attached in the end for final classification. It is noteworthy that we employ full fine-tuning in our experiments, optimizing all layers of Dasheng on the clinical dataset.


\begin{table}[t]
\centering
\begin{minipage}[t]{0.48\textwidth}
\centering
\caption{Model structure for the U-Net model}
\label{tab:U-Net}
\resizebox{\columnwidth}{!}{\normalsize%
\begin{tabular}{|c|c|c|c|}
\hline
Operator  & CH & Kernel & Stride \\ \hline
ConvBlock & 64  & (3,3)  & (1,1)  \\ \hline
MaxPool2D & -   & (2,2)  & -      \\ \hline
ConvBlock & 128 & (3,3)  & (1,1)  \\ \hline
MaxPool2D & -   & (2,2)  & -      \\ \hline
ConvBlock & 256 & (3,3)  & (1,1)  \\ \hline
MaxPool2D & -   & (2,2)  & -      \\ \hline
ConvBlock & 512 & (3,3)  & (1,1)  \\ \hline
UpSample  & -   & -      & -      \\ \hline
ConvBlock & 256 & (3,3)  & (1,1)  \\ \hline
UpSample  & -   & -      & -      \\ \hline
ConvBlock & 128 & (3,3)  & (1,1)  \\ \hline
UpSample  & -   & -      & -      \\ \hline
ConvBlock & 64  & (3,3)  & (1,1)  \\ \hline
ConvBlock & 1   & (1,1)  & (1,1)  \\ \hline
\end{tabular}%
}
\end{minipage}
\hfill
\begin{minipage}[t]{0.48\textwidth}
\centering
\caption{Model structure for the ConvBlock}
\label{tab:convblock}
\resizebox{\columnwidth}{!}{\normalsize%
\begin{tabular}{|c|c|c|}
\hline
Operator    & \# of out channels & Kernel \\ \hline
Conv2D      & CH                 & (3,3)  \\ \hline
BatchNorm2D & CH                 & -      \\ \hline
Conv2D      & CH                 & (3,3)  \\ \hline
BatchNorm2D & CH                 & -      \\ \hline
ReLU        & -                  & -      \\ \hline
\end{tabular}%
}
\end{minipage}
\end{table}

\section{Visual Feature Extraction with Enhanced Glottis Image Segmentation}
\label{sec:glottis_segmentation}
In the evaluation of laryngeal function and pathology, accurate image segmentation of the glottis is essential. This segmentation enables the extraction of objective metrics that clinicians can utilize for diagnosis and review. In our system, we 
implement a two-stage approach comprising a U-Net-based method followed by a diffusion-based refinement. The initial segmentation is performed using a naive U-Net model, which is simple yet effective in medical image segmentation tasks. This model provides a robust initial estimate of the glottis boundaries. 

However, the U-Net model tends to produce false positives~\cite{huang2020unet, vuola2019mask, jaeger2020retina}, especially in cases where the glottal area is not visible~(see the statistics shown in Table~\ref{tab:seg_model}), resulting in segmentation outputs even in the absence of glottal regions. This can significantly impair the accuracy of subsequent analyses, and complicate subsequent vocal fold analysis modules. To address this issue and enhance the precision of the segmentation results, we further refine the U-Net outputs using a diffusion model. This additional step helps to correct the errors and produce a more accurate and reliable glottis mask. The combination of these two methods ensures high-quality segmentation, which is crucial for subsequent analysis and metric computation.

The following subsections provide detailed descriptions of each component in our glottis segmentation pipeline, including the U-Net-based method and the diffusion-based refinement.

\subsection{U-Net-based Method}
\label{subsec:U-Net}
The U-Net model is a convolutional neural network specifically designed for biomedical image segmentation. Owing to its effectiveness and accuracy, we employ a U-Net as our segmentation model. As shown in Table~\ref{tab:U-Net}, the model comprises a series of convolutional blocks, referred to as ConvBlocks (see Table~\ref{tab:convblock}), which include convolution operations followed by batch normalization and ReLU activation functions. The U-Net architecture is symmetrical, featuring a contracting path that captures contextual information and an expansive path that facilitates precise localization through upsampling operations. The contracting path consists of ConvBlocks with progressively increasing numbers of channels (64, 128, 256, and 512), interspersed with MaxPool2D layers for downsampling. This is followed by a sequence of upsampling operations and ConvBlocks that reduce the feature map dimensions, effectively reconstructing the image to its original resolution. The final layer utilizes a ConvBlock with a single output channel to generate the segmentation mask. 

\begin{figure*}[t]
	\centering
	\includegraphics[width=\textwidth]{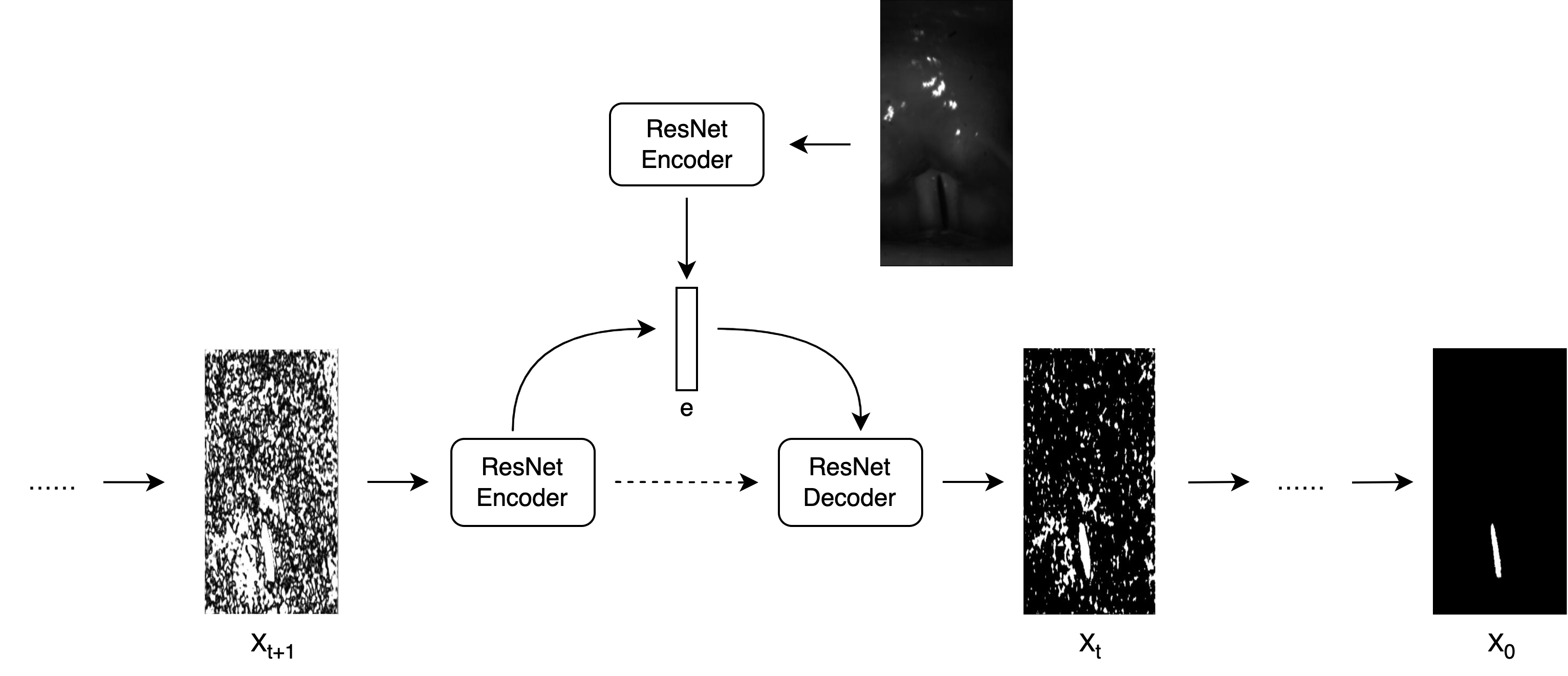}
	\caption{The overview of the second-pass diffusion-based refinement. The denoising process starts from a customized Gaussian noise, utilizing the glottis mask generated by the first-pass U-Net glottis segmentation.}
	\label{fig:diff}
\end{figure*}

\subsection{Diffusion Model-based Refinement}
\label{subsec:diff}
The U-Net model excels at extracting image masks. However, the generated masks may tend to yield false alarms when no glottis area is shown~(see Table~\ref{tab:seg_model}). To further enhance the quality, we explore the integration of the diffusion model. Diffusion models are generative models, which proves to be very powerful tools for generating high-quality images~\cite{ho2020denoising,yang2023diffusion}. These models are designed to generate images by iteratively denoising random noise to match a target distribution, allowing them to produce diverse outputs while still maintaining a coherent structure or theme. By incorporating the diffusion model as a post refinement step, we expect the glottis mask extracted by U-Net to be further optimized with the diversity introduced.

Diffusion models consist of two stages: forward diffusion and reverse diffusion. During the forward process, Gaussian noise is gradually added to the segmentation label \( x_0 \) over a series of steps \( T \). Conversely, in the reverse process, a neural network is trained to recover the original data by reversing the noise addition, as represented by the following equation:
\begin{equation}
    p_\theta\left(x_{0: T-1} \mid x_T\right)=\prod_{t=1}^T p_\theta\left(x_{t-1} \mid x_t\right),
\end{equation}
where \( \theta \) stands for the parameters for the reverse process.

\begin{figure*}[t]
    \centering
    \begin{subfigure}{0.32\textwidth}
        \centering
        \includegraphics[width=\textwidth]{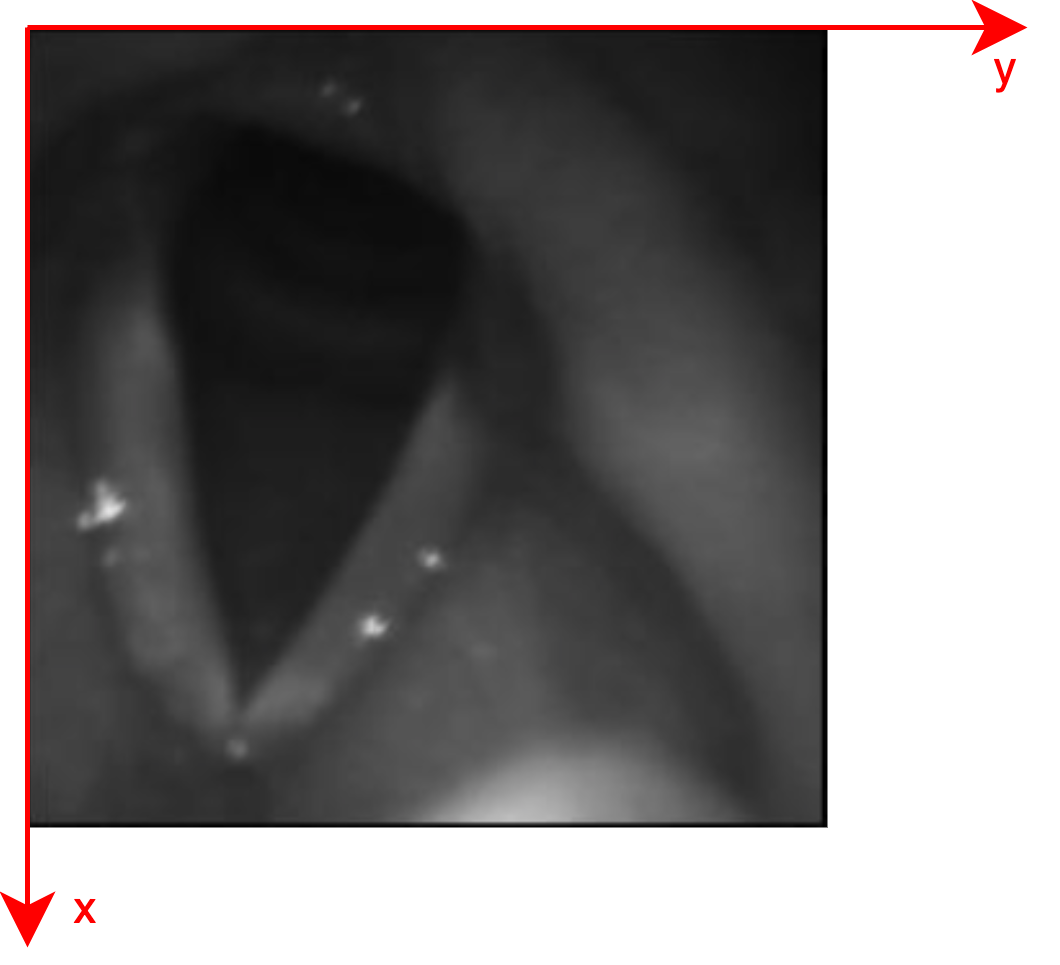}
        \caption{}
    \end{subfigure}
    \begin{subfigure}{0.32\textwidth}
        \centering
        \includegraphics[width=\textwidth]{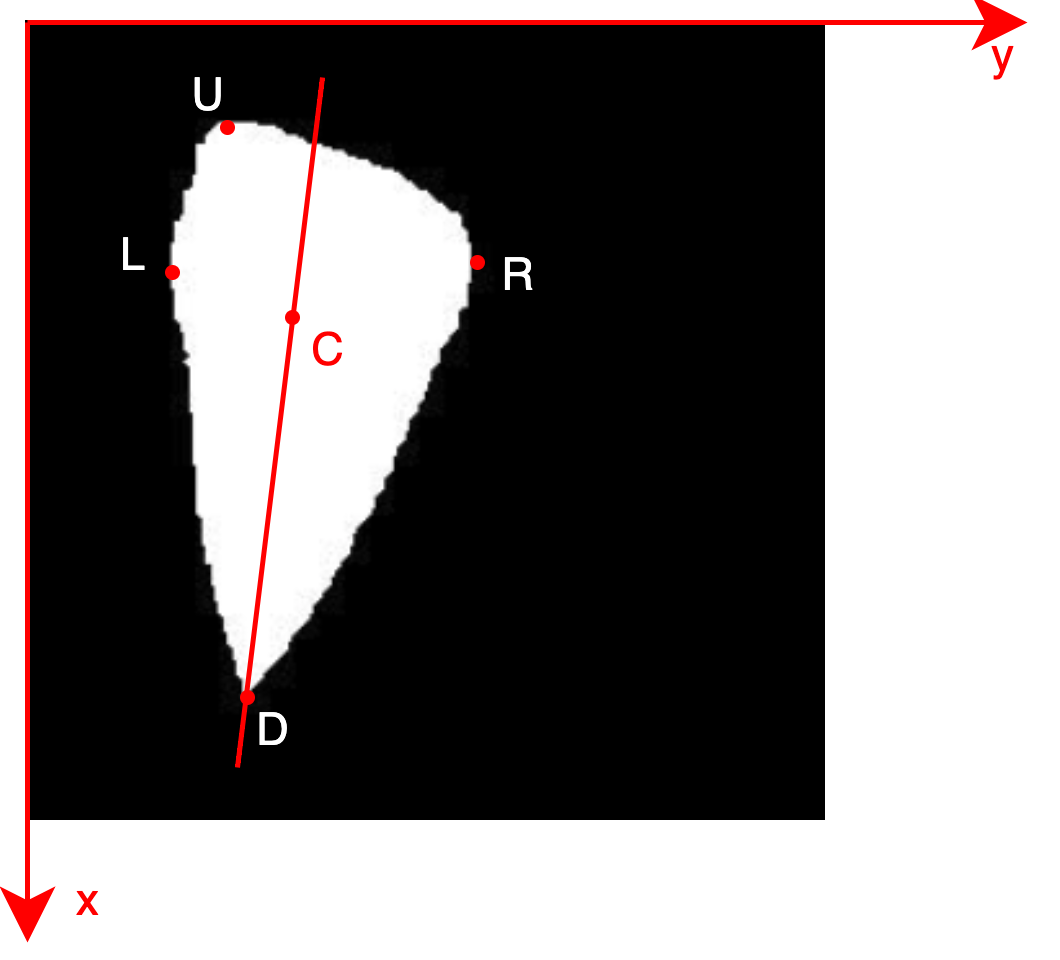}
        \caption{}
    \end{subfigure}
    \begin{subfigure}{0.32\textwidth}
        \centering
        \includegraphics[width=\textwidth]{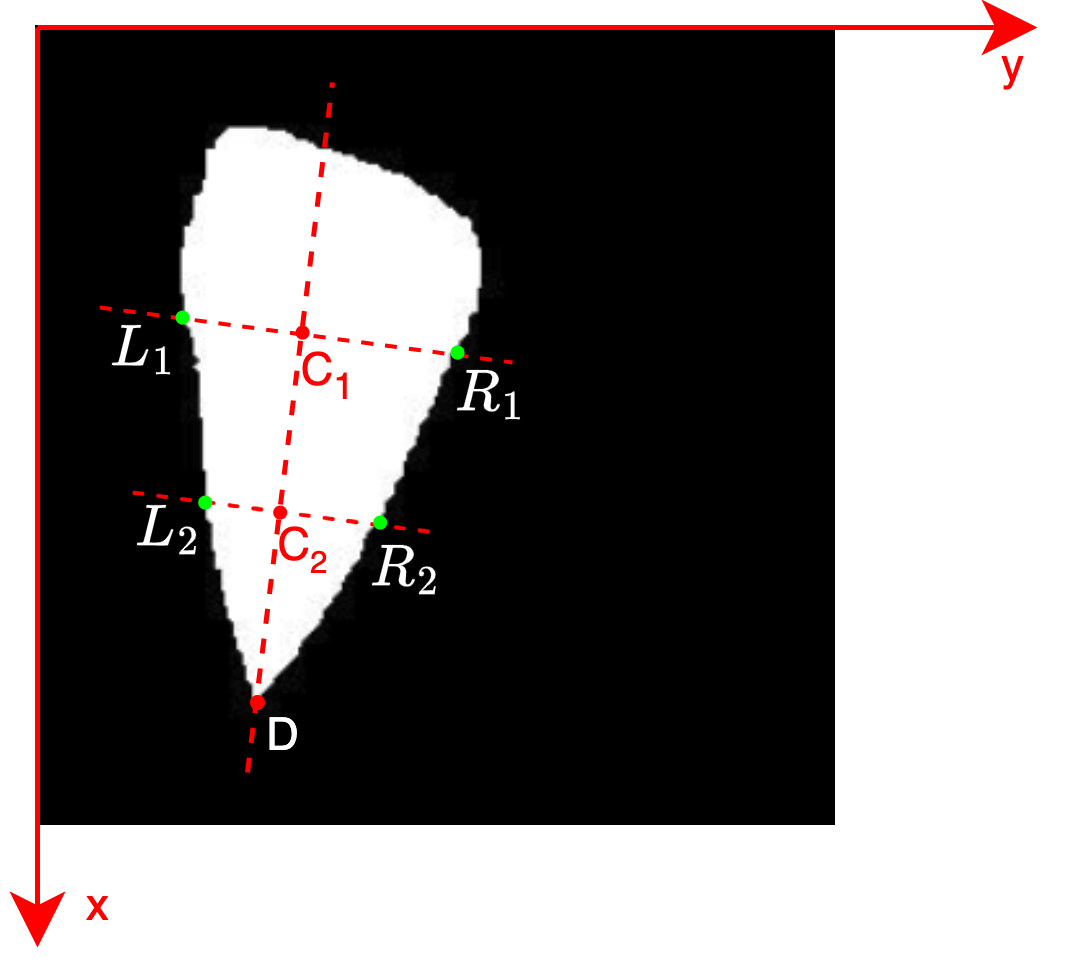}
        \caption{}
    \end{subfigure}
    \\
    \begin{subfigure}{0.32\textwidth}
        \centering
        \includegraphics[width=.95\textwidth]{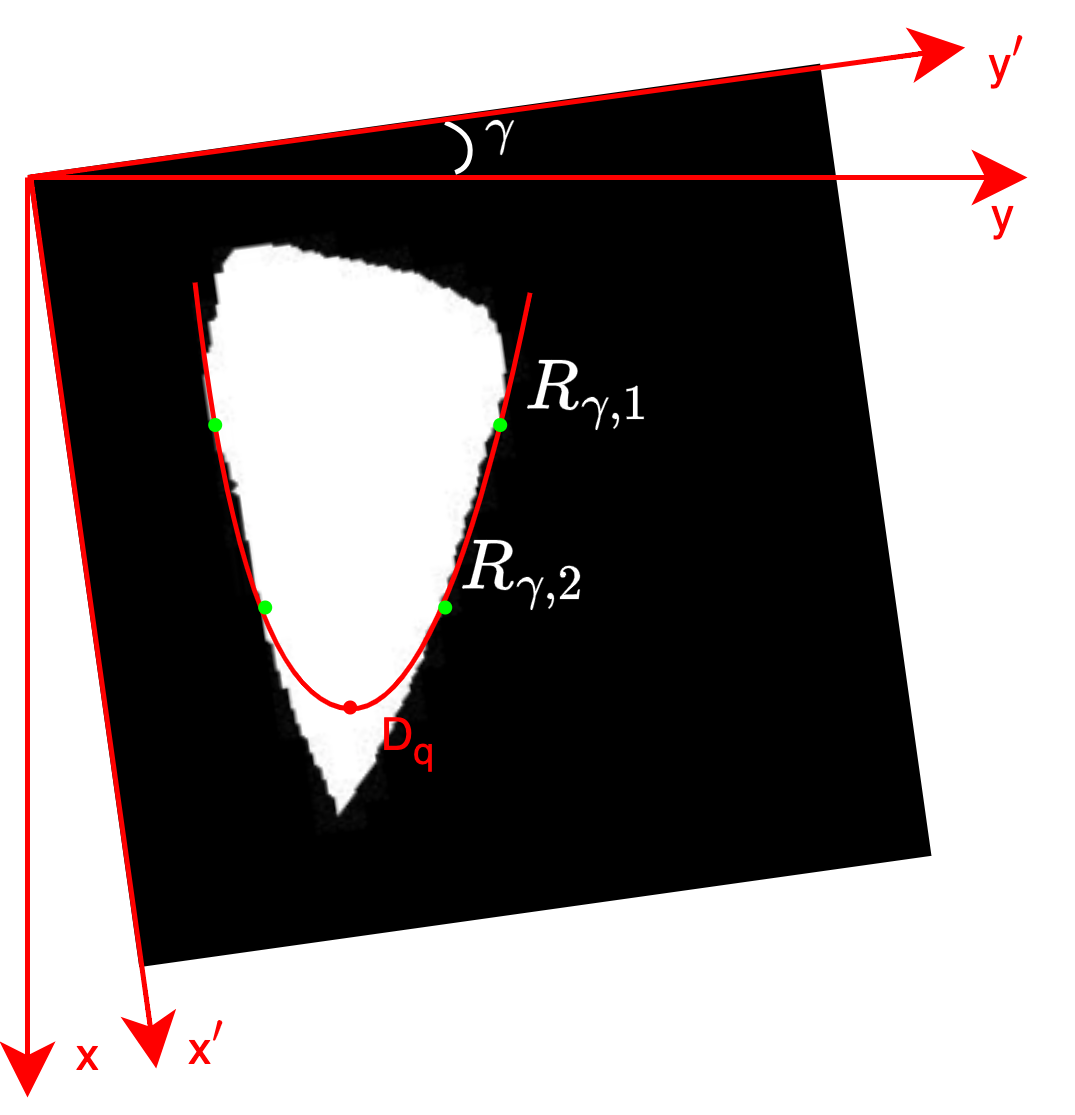}
        \caption{}
    \end{subfigure}
    \begin{subfigure}{0.32\textwidth}
        \centering
        \includegraphics[width=\textwidth]{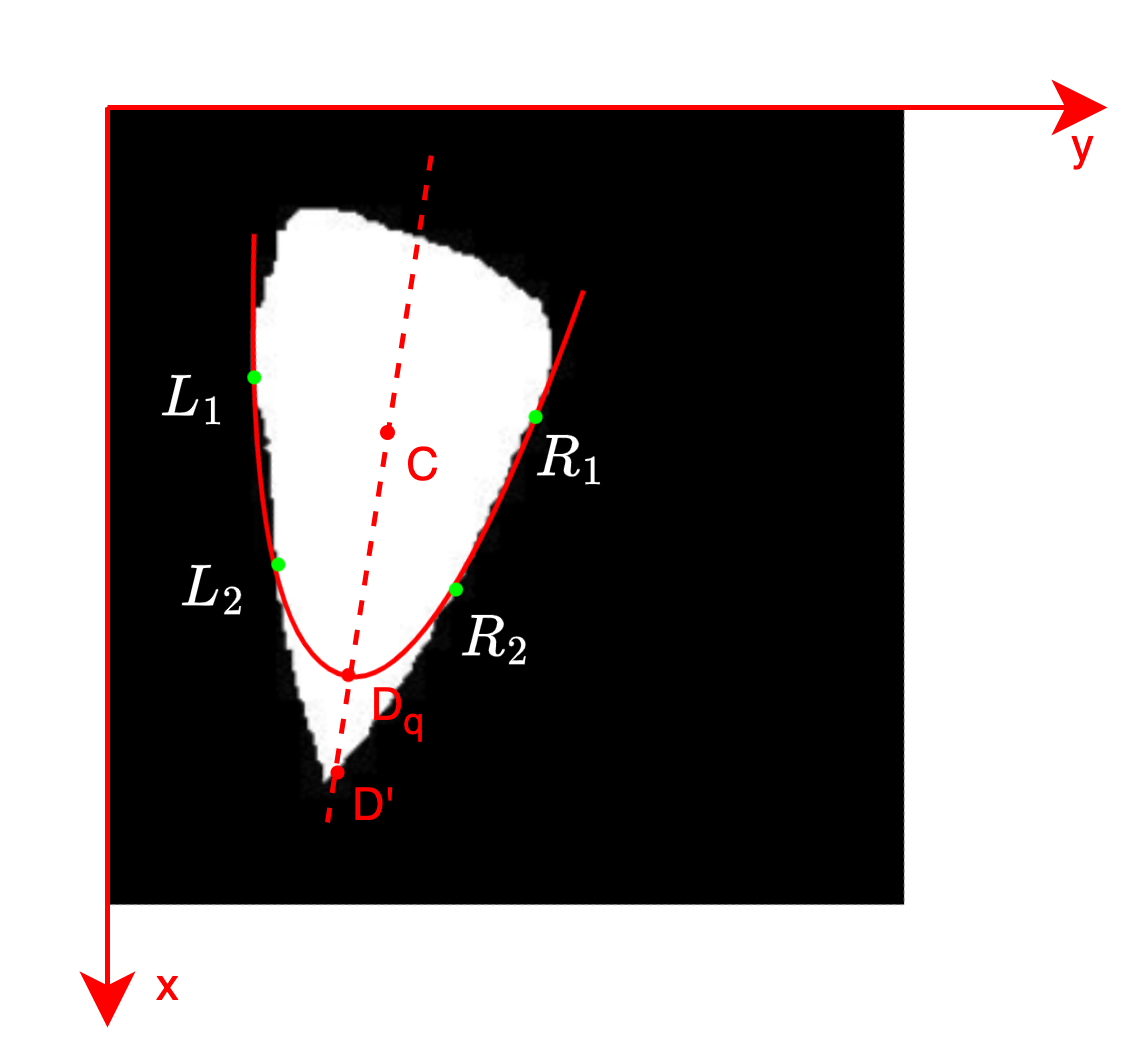}
        \caption{}
    \end{subfigure}
    \begin{subfigure}{0.32\textwidth}
        \centering
        \includegraphics[width=\textwidth]{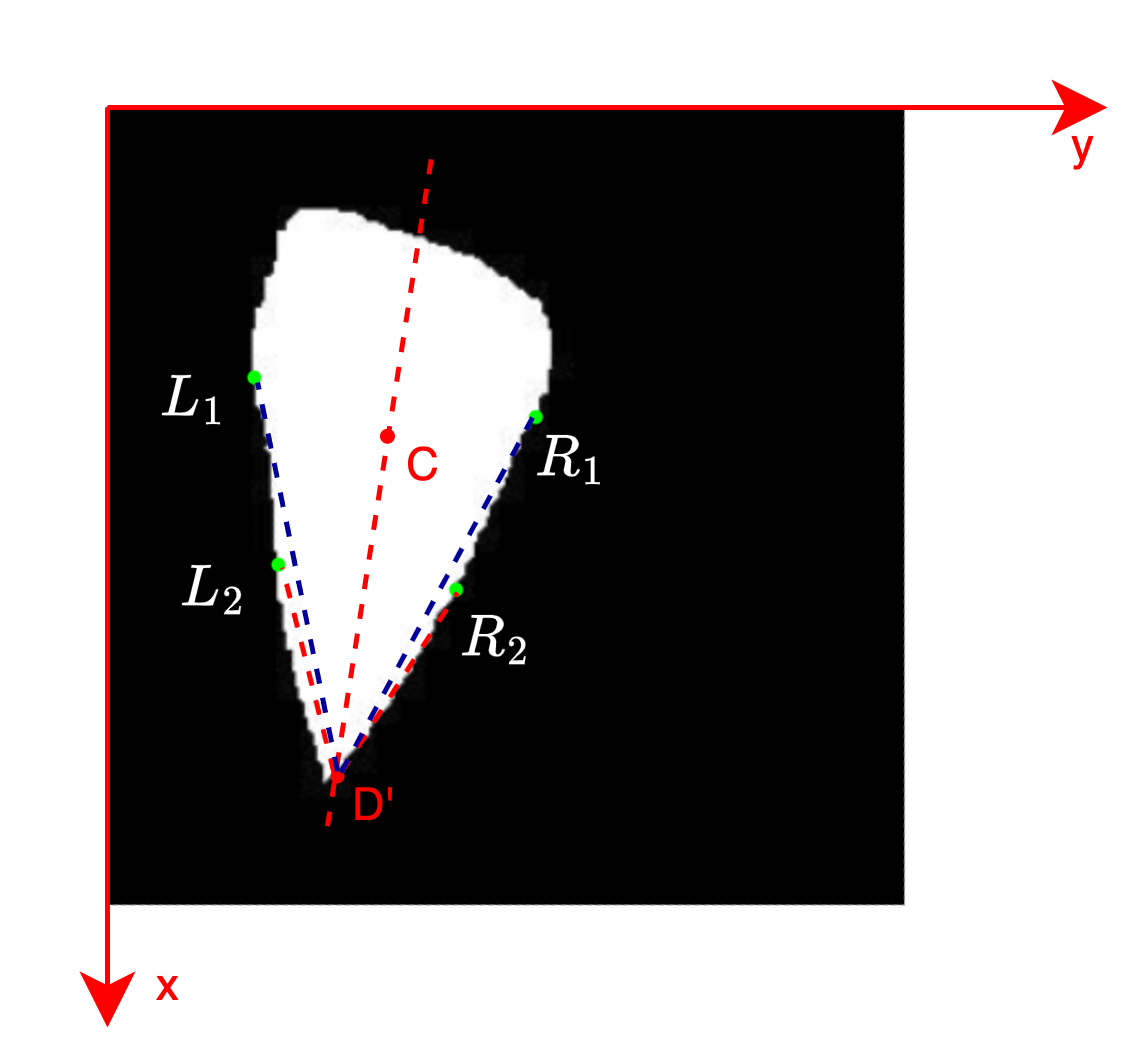}
        \caption{}
    \end{subfigure}
    \caption{The workflow of computing the angle deviation of left and right vocal folds. \added{We use image\#116 from BAGLS dataset as an example. The black background represents the unsegmented region, while the white area corresponds to the segmented glottis mask.} Step~(a) get the center point $C$ and bottom point $D$. Step~(b), (c) and (d) fit the outline of vocal folds with a quadratic function, refining the bottom point to $D^\prime$. As a result, step~(e) shows the refined midline $CD^\prime$. Finally, the angle deviation for both vocal folds can be computed\added{, which are} $\color{black}\angle L_{1}D'C, \angle L_{2}D'C, \angle R_{1}D'C$ \added{and} $\color{black}\angle R_{2}D'C$.}\vspace{-5mm}
    \label{fig:angle_computation}
\end{figure*}

\begin{table}[t]
\centering
\small
\caption{Angle Deviation Extraction Algorithm \added{and Corresponding Figures}.}
\label{tab:angle_extraction}
\begin{tabular}{@{\hskip 8pt}p{0.8\textwidth}p{0.15\textwidth}@{}}
\toprule
\textbf{Algorithm Step} & \textbf{{Figure}} \\ \midrule
1. Compute the center $C(x_c, y_c)$ from $U(x_1, y_1)$, $D(x_2, y_2)$, $L(x_3, y_3)$, $R(x_4, y_4)$. & \textcolor{black}{Figure~\ref{fig:angle_computation}(b)} \\

2. Compute the function $f(x)$ passing through $C$ and $D$. & \textcolor{black}{Figure~\ref{fig:angle_computation}(b)} \\

3. Find $N$ equidistant points $C_1, C_2, \dots, C_{N-1}$ of $f(x)$ on the line segment intercepted by the glottal mask. & \textcolor{black}{Figure~\ref{fig:angle_computation}(c)} \\

4. For each $k = 1, 2, \dots, N-1$: 
\vspace{-2mm}\begin{itemize}\small
    \item[4.1] Compute the function $f_k^\prime(x)$ orthogonal to $f(x)$ passing through $C_k$. 
    \item[4.2] Compute intersection points $L_k$ and $R_k$ between $f_k^\prime(x)$ and the glottis mask.
\end{itemize}\vspace{-6mm}
& \textcolor{black}{Figure~\ref{fig:angle_computation}(c)} \\

5. Rotate the coordinate system by $\gamma$ degrees to align $f(x)$ vertically. & \textcolor{black}{Figure~\ref{fig:angle_computation}(d)} \\

6. Approximate a quadratic curve $q_\gamma(x)$ using the rotated intersection points $\{L_{\gamma, k}\}$ and $\{R_{\gamma, k}\}$. & \textcolor{black}{Figure~\ref{fig:angle_computation}(d)} \\

7. Find the vertex point $D_q$ of $q_\gamma(x)$ and map it back to the original coordinate system. & \textcolor{black}{Figure~\ref{fig:angle_computation}(e)} \\

8. Connect $C$ and $D_q$ to form the corrected midline $f^*(x)$, which intersects the glottis mask at $D^\prime$. & \textcolor{black}{Figure~\ref{fig:angle_computation}(e)} \\

9. Connect $\{L_k\}$ and $\{R_k\}$ with $D^\prime$ to compute the glottal angles $\angle L_k D^\prime C$ and $\angle C D^\prime R_k$. & \textcolor{black}{Figure~\ref{fig:angle_computation}(f)} \\ \bottomrule
\end{tabular}
\end{table}

Consistent with the conventional implementation of the Diffusion Probability Model~(DPM)~\cite{ho2020denoising}, a U-Net model is employed for training. As Figure~\ref{fig:diff} is shown, following the idea of MedSegDiff~\cite{wu2024medsegdiff}, we incorporate the original glottis images as priors for the step estimation function and employ dynamic conditional encoding to fuse the encoding outcomes from both the raw image and the segmentation mask at each step. Hence, for each step, the estimation function \( \epsilon_\theta \) is written as:
\begin{equation}
    \epsilon_\theta\left(x_t, I, t\right)=D\left(\left(E_t^I+E_t^x, t\right), t\right),
\end{equation}
where \( \theta \) stands for the learning parameters, \( D \) is the decoder, \( I \) is the raw image prior, \( t \) is the current diffusion step. \( E_t^I \) and \( E_t^x \) are the embeddings encoded from the raw image and segmentation mask at step \( t \) respectively.

Different from the traditional training procedure, we do not start the diffusion process from a standard Gaussian noise, \( p_\theta(x_T)=\mathcal{N}(x_T; \mathbf{0}, I_{n\times n}) \) for an \( n\times n \) image. Instead, by integrating the U-Net result introduced in Section~\ref{subsec:U-Net}, we start the diffusion process from a customized Gaussian noise, which is:
	\begin{equation}
	    p_\theta(x_T)=\mathcal{N}(x_T; \mu^\prime, I_{n\times n}),
	\end{equation}
	\begin{equation}
	    \mu^\prime = (1 - (\alpha\cdot(1-m^\prime)+(1-\alpha)\cdot m^\prime))\times 10^{-3},
	    \label{eq:mu}
	\end{equation}
where \( \alpha \sim \textit{U}(0,0.3) \) is a random parameter sampled from a uniform distribution, \( m^\prime \) is the mask generated by the U-Net model in Section~\ref{subsec:U-Net}.

We determine the new mean for the Gaussian noise in the diffusion process by computing the weighted average between the glottis mask initially generated by U-Net and its complement~(see Equation~\ref{eq:mu}). This essentially guides the diffusion process to pay more attention to areas outside the glottal region, encouraging the model to refine the segmentation boundaries for improved accuracy.

\subsection{Vocal Fold Dynamics~(VFDyn) Extraction with Quadratic Fitting\added{~(QF)}}
\label{subsec:vfdyn_extraction}
To diagnose the laryngeal paralysis in a more detailed way, instead of AGA or GAW that measures the both left and right vocal folds, we evaluate the angle deviation for each vocal fold. As a result, we can provide a more comprehensive diagnosis of UVFP. The procedure for extracting glottal angles from a segmented glottis involves a systematic series of steps, detailed in \deleted{Algorithm~1}\added{Table~\ref{tab:angle_extraction}}, with visual representations of the intermediate results provided in Figure~\ref{fig:angle_computation}. 

\begin{figure}[t]
	\centering
	\includegraphics[width=\columnwidth]{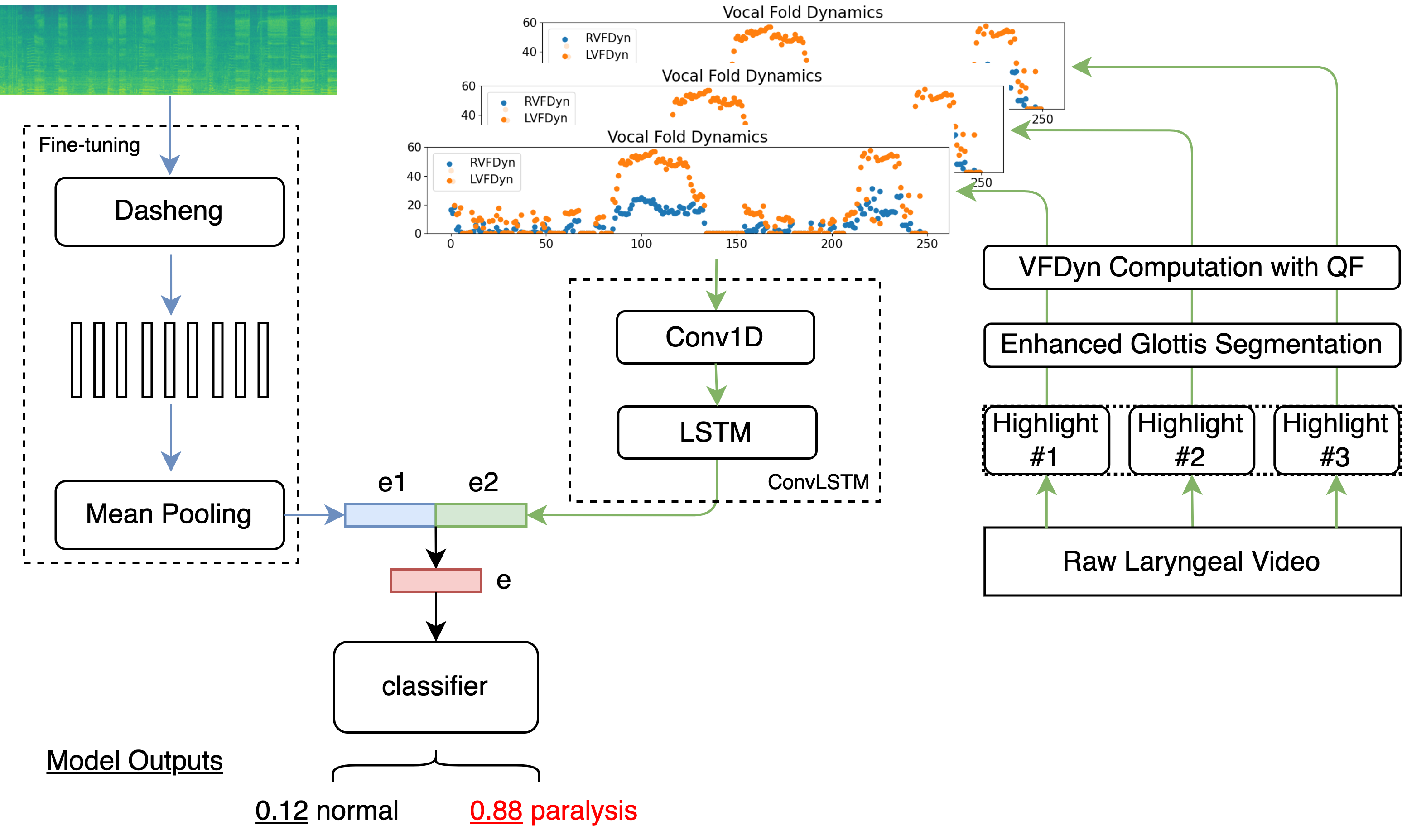}
	\caption{The pipeline of the multimodal VFP classification model.}
	\label{fig:paralysis_detector}
\end{figure}

Initially, the algorithm acquires the coordinates of the top, bottom, left, and right vertices from the glottis mask, computing their center point denoted as \( U(x_1, y_1) \), \( D(x_2, y_2) \), \( L(x_3, y_3) \), \( R(x_4, y_4) \), and \( C(x_c, y_c) \) respectively~(see Figure~\ref{fig:angle_computation}\deleted{(a)}\added{(b)}). Subsequently, a line connecting points \( C \) and \( D \) is established, and a linear function \( f(x) \) passing through \( C \) and \( D \) is computed.

Equidistant points \( C_1, C_2, ..., C_{N-1} \) are then positioned along the line segment intercepted by the glottal mask. For each \( C_k \), orthogonal functions \( f^{\prime}_{k}(x) \) to \( f(x) \) are calculated, with their intersection points \( L_k \) and \( R_k \) with the glottis mask determined for \( k\in [1, N-1] \)~(see Figure~\ref{fig:angle_computation}\deleted{(b)}\added{(c)}).

To ensure uniformity, the coordinate system is rotated by an angle \( \gamma \), aligning all intersection points along the y-axis after rotation, denoted as \( L_{\gamma,k} \) and \( R_{\gamma,k} \). Leveraging these points, a quadratic curve \( q_{\gamma}(x) \) is approximated in the rotated coordinate system~(see Figure~\ref{fig:angle_computation}\deleted{(c)}\added{(d)}. The lowest point \( D_q \) of \( q_{\gamma}(x) \) is identified and mapped back to the original coordinate system~(see Figure~\ref{fig:angle_computation}\deleted{(d)}\added{(e)}).

Subsequently, a calibrated middle line \( f^*(x) \) connecting points \( C \) and \( D_q \) is established, intersecting the glottis mask at point \( D^\prime \). Finally, glottal angles are derived by connecting points \( \{L_k\}_{k\in [1, N-1]} \) and \( \{R_k\}_{k\in [1, N-1]} \) with \( D^\prime \), yielding \( \{\angle L_kD^\prime C\}_{k\in [1, N-1]} \) and \( \{\angle CD^\prime R_k\}_{k\in [1, N-1]} \) respectively\added{~(see Figure~\ref{fig:angle_computation}(f))}.

\section{Multimodal Vocal Fold Paralysis Analysis}
\label{sec:paralysis_analysis}
To detect VFP, we adopt a two-stage process. First, we build a binary classification model to check if it belongs to VFP. Then, with our proposed metrics, left- and right-sided VFP can be identified by comparing the left and right vocal fold movement, which is measured by the extracted LVFDyn and RVFDyn features.

We developed a model that integrates both audio and visual modalities for the enhanced VFP detection. As illustrated in Figure~\ref{fig:paralysis_detector}, the model utilizes the audio Mel spectrograms and the VFDyn as inputs. 
The audio spectrogram is encoded by the Dasheng audio encoder.  
Given that multiple key video segments are extracted from a single laryngoscope video, each associated with corresponding VFDyn features, multiple time series are generated for one video. As described in Section~\ref{subsec:vfdyn_extraction} that multiple equidistant points are selected along the vocal folds, these time series are treated as multi-channel inputs to a lightweight ConvLSTM~\cite{shi2015convolutional} model, which consists of a convolutional layer followed by an LSTM layer. The ConvLSTM model is well-suited for handling multi-channel time series data: the convolutional layer captures spatial features across all channels, while the LSTM layer effectively processes temporal information. The resulting audio embedding $e_1$ and video embedding $e_2$ are then concatenated to form a combined embedding $e$. Finally, a classification layer is added to the model, which outputs the logits for VFP prediction.

To further differentiate between left or right VFP, we compute and compare the variance of the LVFDyn and RVFDyn features. Intuitively, the paralyzed side exhibits less activity during phonation, leading to a smoother time sequence and consequently a lower variance. Therefore, by simply comparing the variance of left and right, we are able to distinguish between left or right VFP.

\section{Experimental Settings}
\label{sec:exp_settings}
\subsection{BAGLS Dataset}
We use the Benchmark for Automatic Glottis Segmentation~(BAGLS)~\cite{gomez2020bagls} as the dataset for evaluating glottis image segmentation. It consists of 59,250 endoscopic glottis images acquired from hundreds of individuals at seven hospitals, which is split into 55,750 training images and 3,500 test images. In this paper, we follow the same data split as~\cite{gomez2020bagls} to train the U-Net model and the diffusion model. Since the dataset only contains segmentation mask labels instead of bounding boxes covering the vocal folds, we utilize those labels to generate bounding boxes for vocal fold object detection.

\subsection{SYSU Dataset}
This dataset was collected in real-world clinical settings by the Sun Yat-sen Memorial Hospital of Sun Yat-sen University (SYSU). This study was approved by the Institutional Review Board of Sun Yat-sen Memorial Hospital of Sun Yat-sen University under no. SYSEC-KY-KS-2022-040 and the Institutional Review Board of Duke Kunshan University under no. 2022ML067, 2023LM161 and 2025LM041.
The dataset is deliberately divided into two parts, SYSU-A and SYSU-B. 

The SYSU-A dataset consists of 520 clinical video samples, including 106 normal cases and 414 paralysis cases~(257 left VFP and 157 right VFP), and is used to evaluate the performance of the proposed MLVAS in UVFP detection. To ensure fairness, in the following experiment, we conduct a stratified 10-fold cross validation on the original dataset, ensuring the ratio between positive and negative samples to be the same for the training and testing dataset.

The SYSU-B dataset contains clinical video samples from cases other than VFP, e.g. vocal fold polyps and vocal fold nodules. We randomly split the SYSU-B dataset into training and testing subsets, assigning 80\% of the samples for training and 20\% for testing. The audio KWS model is trained and validated on the SYSU-B dataset, and its performance is then further evaluated on the SYSU-A dataset. It should be noted that there is no video overlap between the SYSU-A and SYSU-B datasets, ensuring the model can be applied to the full SYSU-A dataset for evaluation.

All videos are recorded using laryngeal videostroboscopes, with each clip containing at least one strobing video segment. The dataset reflects real-world variability, as the recordings were captured using different types of endoscopes across various locations, leading to domain shifts and posing challenges for algorithms to address.


\begin{figure}[t]
    \centering
    \begin{subfigure}{0.48\textwidth}
        \centering
        \includegraphics[width=\textwidth]{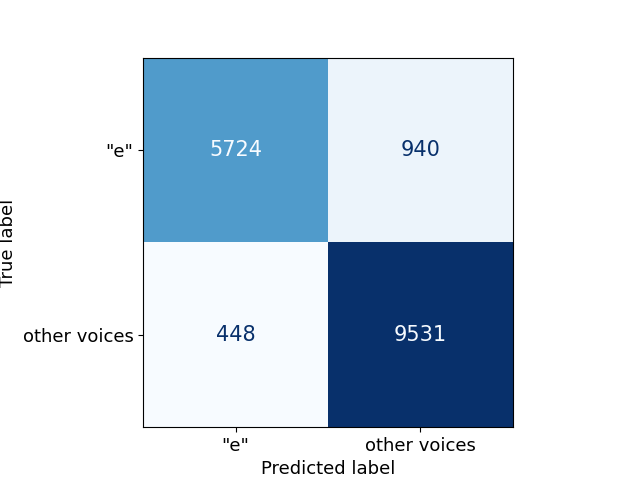}
        \caption{\qquad \qquad SYSU-B-test set}
        \label{fig:cm_kws_train}
    \end{subfigure}
    \begin{subfigure}{0.48\textwidth}
        \centering
        \includegraphics[width=\textwidth]{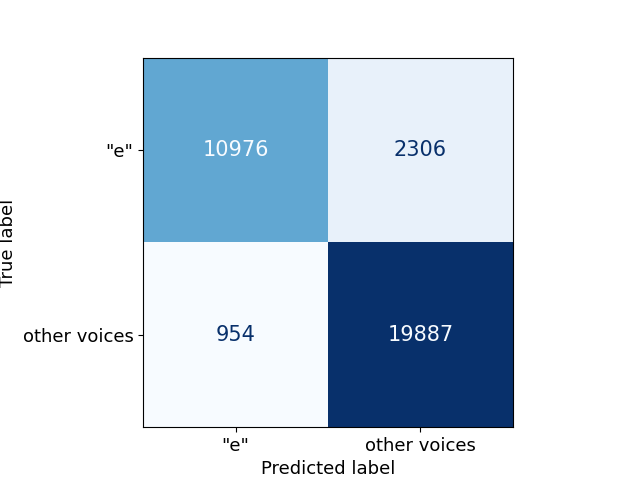}
        \caption{\qquad \qquad SYSU-A set}
        \label{fig:cm_kws_eval}
    \end{subfigure}
    \caption{The confusion matrices of the classification result of the audio KWS model using the best threshold with the highest F-score.}
    \label{fig:cm_kws}
\end{figure}

\begin{table}[t]
\caption{The classification performance~(\%) of the KWS model on the full dataset of SYSU-A and the testing set of SYSU-B with the best threshold.}
\centering
\label{tab:kws_performance}
\resizebox{.8\columnwidth}{!}{\normalsize%
\begin{tabular}{@{}cccccc@{}}
\toprule
        Dataset & \# of samples & Accuracy & Precision & Recall & F-score \\ \midrule
        SYSU-A  & 34123   & 90.45        & 90.54         & 90.45      & 90.34        \\ 
        SYSU-B-test  & 16643   & 91.66        & 91.71         & 91.66      & 91.60        \\ \bottomrule
\end{tabular}%
}
\end{table}

\subsection{Keyword Spotting Model}
The detailed architecture of the KWS model is outlined in Table~\ref{tab:keyword_spotting}. The audio input is represented by a Mel-spectrogram with 80 Mel filters, utilizing 1024 FFT points and a hop length of 512. A sliding window of 400 samples with a hop length of 64 samples is employed to extract frame-level features. During training, audio clips are randomly sampled from the SYSU-B dataset. For each selected audio clip, random chunks of 40 frames are extracted to form batches of training data. The model is trained over 100 epochs using the Adam optimizer with a learning rate of 0.005. During inference, a sliding window of 40 frames with a unit step size is used to process each audio clip in the test dataset.

Since the KWS model outputs a posterior score rather than a direct decision value, a threshold is necessary to determine the final result. 
Consequently, the threshold is selected based on the \added{precision-recall curve on the training portion of the SYSU-B dataset, aiming for the} highest F1-score, a balanced metric that considers both precision and recall, providing a comprehensive measure of classification performance. \deleted{In MLVAS}\added{As a result}, we pick a threshold of 0.38 as the classification boundary. If the posterior score output by the KWS model is higher than this value, the input chunk is regarded as \deleted{"e"}\added{``/\textepsilon:/''}. Otherwise, it is treated as other voices. Figure~\ref{fig:cm_kws_train} and Figure~\ref{fig:cm_kws_eval} display the confusion matrix computed on the SYSU-B-test and SYSU-A dataset respectively. The accuracy, precision, recall and F-score are listed in Table~\ref{tab:kws_performance}. As is shown in Table~\ref{tab:kws_performance}, the performance of our audio KWS model are all above 90\% on both the SYSU-A and SYSU-B.

\begin{table}[t]
\centering
\caption{Glottis Image Segmentation Performance on the Testing Set of the BAGLS dataset.}
\label{tab:seg_model}
\resizebox{.5\columnwidth}{!}{\normalsize%
\begin{tabular}{@{}ccc@{}}
\toprule
                                                                        & IoU  & FAR     \\ \midrule
U-Net Baseline                                                          & 0.77 & 15.8\% \\
Diffusion-based                                                         & 0.78 & 4.5\%  \\
U-Net w. diffusion refinement & 0.80 & 2.0\%  \\ \bottomrule
\end{tabular}%
}
\end{table}

\subsection{Glottis Segmentation}
\label{subsec:glottis_segmentation}
We train the U-Net baseline with the same strategy as~\cite{gomez2020bagls}. For the diffusion model, it is built on top of the model in~\cite{wu2024medsegdiff}, except that we slightly modify the noise to meet the prior knowledge of U-Net results. The diffusion step is set to 1,000. We train the diffusion model with 100,000 steps. The model is optimized by a AdamW optimizer with a learning rate of 0.0001.

We evaluate our segmentation model on the public glottis dataset BAGLS. We use Intersection over Union~(IoU) as the metric, and the results are shown in Table~\ref{tab:seg_model}. From the table, we can see that the diffusion model can have a slightly better IoU performance than the traditional U-Net, By refining the U-Net results with the diffusion ones, the IoU performance can be further improved. 

Besides IoU, we  also compute the False Alarm Rate~(FAR) when using different segmentation methods. The False Alarm~(FA) is defined as the situation where the glottis mask is detected when actually there is no glottis in the image. The FAR is calculated as the number of FA divided by the total number of images with no glottis. We care about FA as we need the segmentation masks for later analysis. FAs might produce wrong measurements and mislead the prediction results of VFP. From Table~\ref{tab:seg_model}, the traditional U-Net model shows a tendency to produce FA masks with a leading FAR of 15.8\%. By contrast, diffusion-based segmentation model has a much lower FAR of 4.5\%, which is more than two times lower. The result may indicates that, in terms of the glottis segmentation, the diffusion model performs more robust when dealing with image with no targets. Hence, after we refine the U-Net result with the diffusion model, we can further achieve a result with a lower FAR but a better IoU.

\subsection{Vocal Fold Paralysis Detection}
\label{subsec:vfp_settings}
The multimodal VFP detection model consider both audio and video modalities. The input for audio modality is 64-dimensional log-Mel spectrogram, extracted every 10 ms with a window size of 32 ms for clips of length 10 s. RVFDyn and LVFDyn are used for the video modalities. As for the models, we use Dasheng-Base that has a model size of 86 M parameters as the backbone pre-trained model for audio modeling. ConvLSTM (described in Section~\ref{sec:paralysis_analysis}) is used for video modeling.

During training, we adopt a full fine-tuning strategy with the learning rate set to $5\times10^{-5}$. The models are trained for 800 steps using a batch size of 12, comprising 8 positive samples and 4 negative samples. As each laryngeal video typically contains 2--3 key video segments, one segment is randomly selected from each video for training, ensuring that each batch includes video segments from 12 unique laryngeal videos. During inference, class logits are computed for each key segment within a laryngeal video, and the final logits for the video are obtained by averaging the logits across all its key segments.

\subsection{Evaluation Metrics}
To comprehensively evaluate the performance of our proposed system, we employed several widely recognized metrics, including accuracy, precision, sensitivity~(recall), specificity, F-score, the mean of sensitivity and specificity, and the area under the receiver operating characteristic curve (ROC-AUC). Together, these metrics\deleted{, as detailed in Table~7,} highlight the robustness and clinical applicability of the proposed system.

\begin{sidewaystable}
\caption{The average classification performance (\%) achieved using different audio-based methods, along with the standard deviation~(reported in parentheses). \added{The results are} \deleted{is} evaluated \deleted{across experiments} using 10-fold cross-validation on the SYSU-A dataset. \added{All models operate on pre-extracted key video segments, which ensures the visibility of the larynx and includes at least one complete phonatory cycle.}}
\label{tab:audio-model}
\resizebox{\textwidth}{!}{\Large %
\begin{tabular}{@{}cllccccccccc@{}}
\toprule
\multicolumn{3}{c}{Model} &
  Feature &
  ROC-AUC &
  Accuracy &
  Precision &
  Recall &
  F-score &
  Sensitivity &
  Specificity &
  \begin{tabular}[c]{@{}c@{}}Mean of \\ Sensi. \& Speci.\end{tabular} \\ \midrule
\multicolumn{3}{c}{RF~\cite{cesarini2024multi}} &
  MFCC &
  78.59($\pm$7.84) &
  56.23($\pm$6.42) &
  53.68($\pm$4.17) &
  \textbf{99.27($\pm$1.56)} &
  69.56($\pm$3.10) &
  \textbf{99.27($\pm$1.56)} &
  13.18($\pm$13.87) &
  56.23($\pm$6.42) \\
\multicolumn{3}{c}{SVM} &
  MFCC &
  74.81($\pm$7.56) &
  68.41($\pm$6.75) &
  70.88($\pm$8.71) &
  64.72($\pm$6.78) &
  67.28($\pm$5.96) &
  64.72($\pm$6.78) &
  72.09($\pm$12.36) &
  68.41($\pm$6.75) \\
\multicolumn{3}{c}{NB} &
  MFCC &
  75.63($\pm$7.86) &
  71.29($\pm$6.93) &
  71.35($\pm$7.62) &
  72.67($\pm$8.28) &
  71.68($\pm$6.37) &
  72.67($\pm$8.28) &
  69.91($\pm$11.27) &
  71.29($\pm$6.93) \\
\multicolumn{3}{c}{MLP} &
  MFCC &
  72.21($\pm$7.85)&
  60.72($\pm$8.99)&
  59.08($\pm$8.85)&
  83.53($\pm$11.27)&
  68.17($\pm$5.40)&
  83.53($\pm$11.27)&
  37.91($\pm$24.21)&
  60.72($\pm$8.99)\\ \midrule
\multicolumn{3}{c}{ResNet50~\cite{kim2024classification}} &
  MFCC &
  75.09($\pm$6.02)&
  68.74($\pm$6.16)&
  74.76($\pm$14.74)&
  64.48($\pm$15.81)&
  66.42($\pm$8.81)&
  64.48($\pm$15.81)&
  73.00($\pm$16.46)&
  68.74($\pm$6.16)\\
\multicolumn{3}{c}{MobilenetV3} &
  Spec. &
  78.75($\pm$5.64)&
  72.59($\pm$5.10)&
  \textbf{81.34($\pm$9.68)}&
  61.09($\pm$13.02)&
  68.39($\pm$7.74)&
  61.09($\pm$13.02)&
  \textbf{84.09($\pm$11.02)}&
  72.59($\pm$5.10)\\
\multicolumn{3}{c}{Dasheng} &
  Spec. &
  \textbf{85.47($\pm$6.37)}&
  \textbf{75.18($\pm$8.42)}&
  71.17($\pm$8.58)&
  88.63($\pm$5.34)&
  \textbf{78.49($\pm$5.40)}&
  88.63($\pm$5.34)&
  61.73($\pm$18.34)&
  \textbf{75.18($\pm$8.42)}\\ \bottomrule
\end{tabular}%
}

\bigskip\bigskip
\caption{The average classification performance (\%) achieved using different enhanced modules and modalities, along with the standard deviation~(reported in parentheses), is evaluated across experiments using 10-fold cross-validation on the SYSU-A dataset.\added{~Audio: Dasheng-encoded audio features; VFDyn: Vocal fold dynamics extracted from key video segments.}}
\label{tab:paralysis}
\resizebox{\textwidth}{!}{\Huge%
\begin{tabular}{@{}ccccccccccc@{}}
\toprule
Modality &
  \begin{tabular}[c]{@{}c@{}}Quadratic-\\ -fitting\end{tabular} &
  \begin{tabular}[c]{@{}c@{}}Diffusion-\\ -refinement\end{tabular} &
  ROC-AUC &
  Accuracy &
  Precision &
  Recall &
  F-score &
  Sensitivity &
  Specificity &
  \begin{tabular}[c]{@{}c@{}}Mean of\\ Sensi. \& Speci.\end{tabular} \\ \midrule
Audio &
  - &
  - &
  85.47($\pm$6.37)&
  75.18($\pm$8.42)&
  71.17($\pm$8.58)&
  88.63($\pm$5.34)&
  78.49($\pm$5.40)&
  88.63($\pm$5.34)&
  61.73($\pm$18.34)&
  75.18($\pm$8.42) \\
Audio+\deleted{AGA}\added{VFDyn} &
  - &
  - &
  84.88($\pm$7.66) &
  75.57($\pm$9.57) &
  71.73($\pm$9.49) &
  87.41($\pm$8.14) &
  78.42($\pm$7.64) &
  87.41($\pm$8.14) &
  63.73($\pm$16.92) &
  75.57($\pm$9.57) \\
Audio+\deleted{AGA}\added{VFDyn} &
  \checkmark &
  - &
  85.82($\pm$4.71) &
  76.20($\pm$8.86) &
  \textbf{73.67($\pm$9.92)} &
  85.03($\pm$6.33) &
  78.49($\pm$6.57) &
  85.03($\pm$6.33) &
  \textbf{67.37($\pm$17.47)} &
  76.20($\pm$8.86) \\
Audio+\deleted{AGA}\added{VFDyn} &
  \checkmark &
  \checkmark &
  \textbf{87.04($\pm$3.69)} &
  \textbf{78.12($\pm$5.06)} &
  73.32($\pm$6.03) &
  \textbf{90.06($\pm$6.52)} &
  \textbf{80.52($\pm$4.00)} &
  \textbf{90.06($\pm$6.52)} &
  66.18($\pm$11.35) &
  \textbf{78.12($\pm$5.06)} \\ \bottomrule
\end{tabular}%
}
\end{sidewaystable}

\section{Experimental Results}
\label{sec:exp_results}

\subsection{Detection of Vocal Fold Paralysis}
\label{subsec:vfp_performance}
To evaluate the effectiveness of our proposed methods, we conduct ablation studies using various system configurations on the clinical dataset. A stratified 10-fold cross-validation approach is employed to ensure robust evaluation. The results are reported as the mean and standard deviation across the 10 folds. The experiments primarily focus on three key aspects: 
\begin{enumerate}
    \item \textbf{Effectiveness of the pre-trained audio model:} Assessing its impact on VFP detection performance comparing with popular SOTA methods.
    \item \textbf{Impact of metric-enhancing modules:} Evaluating the contributions of the \deleted{Quadratic Fitting~(}QF\deleted{)} and Diffusion Refinement~(DR) modules in improving metric extraction, and thus improving the VFP detection performance.
    \item \textbf{Multimodality in VFP detection:} Analyzing the benefit of integrating audio and video modalities for enhanced detection performance.
\end{enumerate}
\subsubsection{Audio Models vs. VFP Detection Performance}

We evaluate the effectiveness of employing pre-trained audio models for VFP prediction by comparing our proposed solution with other widely adopted methods in the field. The results, summarized in Table~\ref{tab:audio-model}, include RF with MFCC, a model shown to achieve SOTA performance on vocal pathology classification tasks~\cite{cesarini2024multi}. ResNet50 with MFCC, which has demonstrated superior performance over traditional machine learning models for VFP detection on other datasets~\cite{kim2024classification}, is also included. Additionally, we consider commonly used baselines, including SVM, NB, and MLP~\cite{cesarini2024multi, kim2024classification}. For deep learning comparisons, MobileNetV3~\cite{howard2019searching} is evaluated due to its proven efficiency and effectiveness in image classification tasks.

Our method demonstrates a superior overall balance across these metrics. This is evident from its remarkable F-score and the mean of sensitivity and specificity. Notably, our proposed method achieves an F-score of 78.49\%, which outperforms the second-best model by nearly 7\%. Furthermore, it maintains a balanced performance in sensitivity and specificity, achieving values of 88.63\% and 61.73\%, respectively. This results in the highest mean of sensitivity and specificity at 75.18\%. Compared to other models, our method achieves high sensitivity while maintaining relatively low specificity, which is crucial for MLVAS in assisting clinicians. This balance reduces the risk of missing unhealthy patients while avoiding the misdiagnosis of an excessive number of healthy patients.

In terms of sensitivity (recall), as shown in Table~\ref{tab:audio-model}, RF with MFCC achieves an impressive value of 99.27\%. However, this is accompanied by a significantly lower precision (53.68\%), suggesting a tendency to over-predict positive labels and generate a high number of false alarms. Conversely, MobileNetV3 attains the highest precision of 81.34\%, but its sensitivity is relatively low at 61.09\%, indicating limited capability in detecting positive cases and potentially overlooking numerous patients genuinely affected by VFP.

Our method leverages fine-tuning of a pre-trained model to achieve an effective balance between precision and recall, thus addressing the limitations observed in other models. This balance underscores the robustness of Dasheng across all performance metrics, positioning it as a reliable and practical approach for VFP detection.

\subsubsection{Enhanced Modules vs. VFP Detection Performance}

Accurate midline prediction is critical for analyzing left and right vocal fold movements, as it serves as a reference for measuring angle deviations and calculating distances between sample points on the vocal folds and the midline. However, in real-world applications, the laryngeal camera’s position is not always fixed, complicating metric computation. Calibrating the midline by fitting a quadratic function has proven effective for improving midline prediction, leading to better classification performance. The associated results are detailed in Table~\ref{tab:paralysis}. The system's native setting serves as a foundational benchmark from which incremental enhancements are realized. The introduction of the proposed QF method improves the specificity from 63.73\% to 67.37\%, reducing the risk of misdiagnosing healthy patients. Further improvements are achieved by incorporating DR into the glottis segmentation process. As indicated in Table~\ref{tab:paralysis}, all the metrics show great improvement, except precision and specificity show a slight decrease. ROC-AUC, accuracy, and F-score improve by approximately 2\% in absolute terms. Particularly, sensitivity grows from 85.03\% to 90.06\%, significantly reducing the probability of missing a negative sample. DR is applied only to images where the glottis is not detected by the YOLO-v5 detector~(discussed in Section~\ref{subsec:video_processing_module}), as this technique specifically targets false alarms~(FAs) in glottis detection~(described in Section~\ref{subsec:glottis_segmentation}). By reducing false alarms, the overall performance of downstream VFP detection improves.


\subsubsection{Multimodality vs. VFP Detection Performance}

We also investigated the effect of multimodal inputs on VFP detection. When the AGA movement metrics were removed, classification performance decreased on all the metrics, as shown in Table~\ref{tab:paralysis}. Specifically, specificity fall by an absolute value of 4.45\%, and all the other metrics drop by an approximate value of 2\% in absolute terms. These results highlight the importance of incorporating multimodality for robust performance. 

\begin{table}[t]\color{black}
\centering
\caption{\textcolor{black}{One-tailed paired $t$-test $p$-values ($\alpha = 0.05$) comparing the proposed system against ablated and unimodal baselines across key performance metrics. Asterisks (*) denote statistically significant improvements. DR: Diffusion-based refinement. QF: Quadratic fitting. Audio-only: System using only audio modality without visual features.}}
\resizebox{\textwidth}{!}{%
\begin{tabular}{lcccc}
\toprule
{Proposed System} & {Precision} & {Specificity} & {F1-score} & {Mean(Sens, Spec)}  \\
\midrule
{vs. w/o DR}        & {0.019566$^*$} & {0.011240$^*$} & {0.005401$^*$} & {0.004115$^*$} \\
{vs. w/o DR or QF}  & {0.027885$^*$} & {0.006049$^*$} & {0.033392$^*$} & {0.009962$^*$} \\
{vs. Audio-only}    & {0.004002$^*$} & {0.001061$^*$} & {0.003680$^*$} & {0.001114$^*$} \\
\bottomrule
\end{tabular}
}
\label{tab:ttest_results}
\end{table}

\textcolor{black}{\subsubsection{Statistical Significance}}

\added{To further validate the effectiveness of our proposed system—particularly the impact of the enhancement modules and multimodal integration—we conduct statistical significance testing using paired $t$-tests. Given the relatively small number of folds ($n=10$) in the initial cross-validation results reported in Table~\ref{tab:paralysis}, applying parametric tests directly on these samples would lack sufficient statistical power. To address this, we adopt a 10$\times$ repeated stratified 10-fold cross-validation scheme, generating 100 runs per method. This approach increases statistical robustness and aligns with best practices recommended in prior work~\cite{wong2019reliable,rodriguez2009sensitivity}.}

\added{As reported in Table~\ref{tab:ttest_results}, all $p$-values fall below the one-tailed significance threshold of $\alpha = 0.05$, indicating that the proposed system achieves statistically significant improvements over the baselines across all evaluation metrics. The first two comparisons isolate the contributions of our enhancement modules. The significant gains confirm that the proposed refinements—quadratic midline fitting and diffusion-based segmentation—substantially improve feature quality and detection accuracy. The third comparison, labeled ``Audio-only,'' evaluates the added value of incorporating visual features (VFDyn) into the system. The statistically significant improvement further demonstrates the importance of visual information for achieving robust and reliable VFP detection.}





\begin{figure*}[t]
    \centering
    \begin{minipage}[b]{0.65\textwidth}
	    \centering
	    \begin{subfigure}{0.32\textwidth}
	        \centering
	        \includegraphics[width=\textwidth]{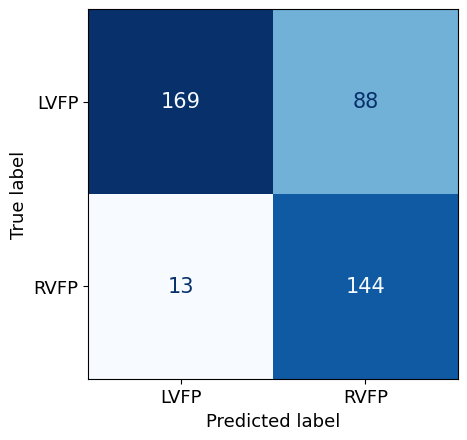}
	        \caption{System 1}
	    \end{subfigure}
	    \begin{subfigure}{0.32\textwidth}
	        \centering
	        \includegraphics[width=\textwidth]{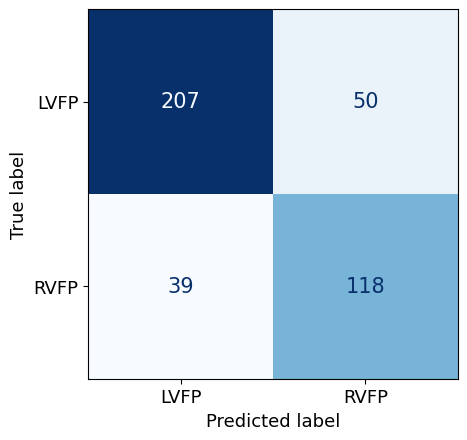}
	        \caption{System 2}
	    \end{subfigure}
	    \begin{subfigure}{0.32\textwidth}
	        \centering
	        \includegraphics[width=\textwidth]{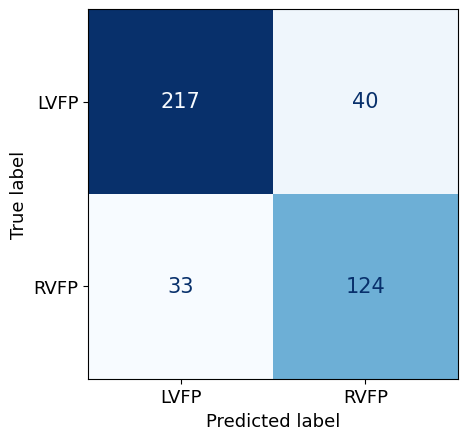}
	        \caption{Proposed System}
	    \end{subfigure}
	    \caption{Confusion matrix of the proposed system for the left or right VFP classification on the VFP cases of the SYSU-A dataset. System 1: w/o QF and DR. System 2: w/o DR. The proposed system: w/ QF and DR.}
	    \label{fig:cm_uvfp}
    \end{minipage}
    \hfill
    \begin{minipage}[b]{0.33\textwidth}
	    \centering
        \includegraphics[width=.9\textwidth]{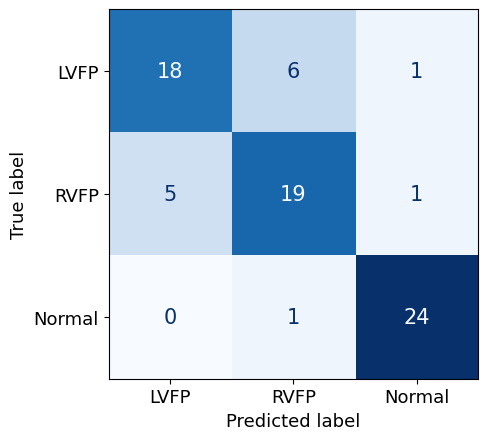}
	    \caption{Confusion matrix of the proposed system for \deleted{the overall subject-level} UVFP classification on the SYSU-A test dataset.}
	    \label{fig:cm_uvfp_pipe}
    \end{minipage}
\end{figure*}

\begin{figure*}[t]
    \centering
    \begin{subfigure}{0.49\textwidth}
        \centering
        \includegraphics[width=\textwidth]{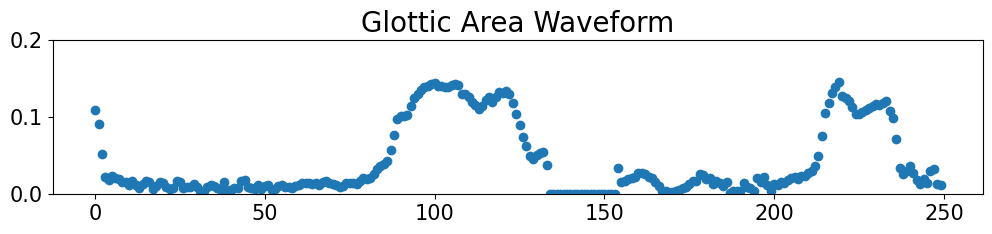}
        \label{fig:gaw_1}
    \end{subfigure}
    \begin{subfigure}{0.49\textwidth}
        \centering
        \includegraphics[width=\textwidth]{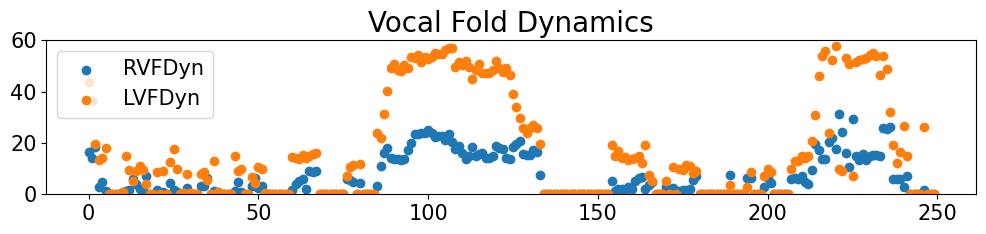}
        \label{fig:vfdyn_1}
    \end{subfigure}
    \begin{subfigure}{0.49\textwidth}
        \centering
        \includegraphics[width=\textwidth]{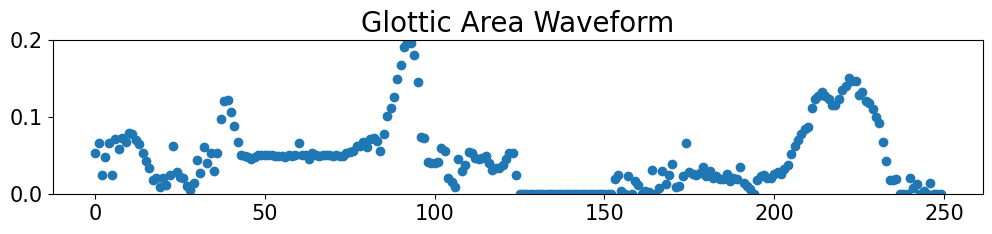}
        \label{fig:gaw_2}
    \end{subfigure}
    \begin{subfigure}{0.49\textwidth}
        \centering
        \includegraphics[width=\textwidth]{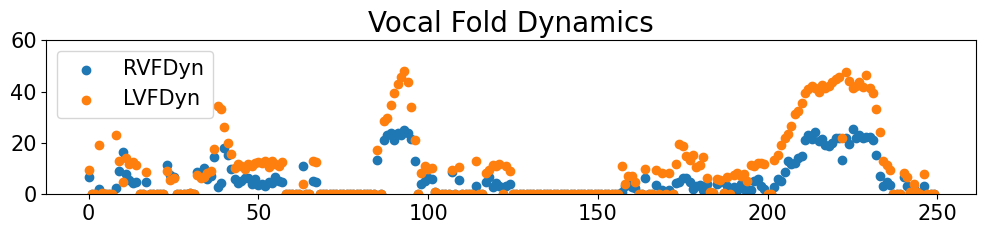}
        \label{fig:vfdyn_2}
    \end{subfigure}
    \begin{subfigure}{0.49\textwidth}
        \centering
        \includegraphics[width=\textwidth]{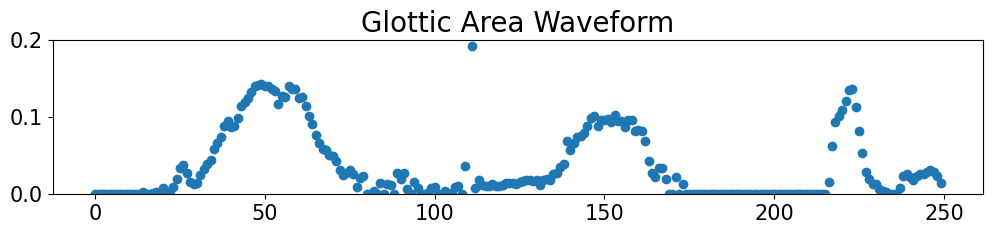}
        \label{fig:gaw_3}
    \end{subfigure}
    \begin{subfigure}{0.49\textwidth}
        \centering
        \includegraphics[width=\textwidth]{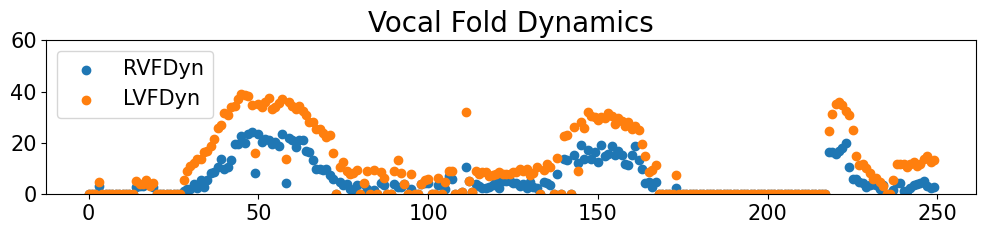}
        \label{fig:vfdyn_3}
    \end{subfigure}
    \caption{The example charts of the extracted GAW and VFDyn from MLVAS for patient\#7530 in the SYSU-A dataset, diagnosed as right VFP. There are three video highlights automatically extracted by the MLVAS~(listed vertically). Thus, three different pairs of charts are extracted, with each pair contains a GAW chart and VFDyn chart listed horizontally.}
    \label{fig:analysis_results}
\end{figure*}

\begin{table}[t]
\centering
\caption{The classification performance (\%) of unilateral vocal fold paralysis classification through VFDyn comparison between left and right vocal folds, extracted by different enhancement modules.}
\label{tab:UVFP}
\resizebox{\columnwidth}{!}{\normalsize %
\begin{tabular}{cccccccc}
\hline
\begin{tabular}[c]{@{}c@{}}Quadratic-\\ -fitting\end{tabular} &
  \begin{tabular}[c]{@{}c@{}}Diffusion-\\ -refinement\end{tabular} &
  Accuracy &
  Precision &
  Recall &
  F-score &
  Sensitivity &
  Specificity \\ \hline
-          & -          & 75.60          & 81.18          & 75.60          & 75.87          & 75.60          & 75.59          \\
\checkmark & -          & 78.50          & 78.87          & 78.50          & 78.63          & 78.50          & 78.55          \\
\checkmark & \checkmark & \textbf{82.37} & \textbf{82.56} & \textbf{82.37} & \textbf{82.44} & \textbf{82.37} & \textbf{82.35} \\ \hline
\end{tabular}
}
\end{table}

\subsection{Detection of Unilateral Vocal Fold Paralysis}
Through the analysis of variance in LVFDyn and RVFDyn, our approach can differentiate between left and right VFP within laryngeal stroboscopic videos. The performance of our detection system for UVFP is initially demonstrated on the SYSU-A dataset, where it effectively distinguishes between left and right VFP instances. Subsequently, the system's broader performance is evaluated across the entire dataset with the VFP detection results from the first-stage multimodal VFP analysis, providing a comprehensive view of MLVAS's real-world clinical applicability.

The classification results on the VFP cases from the SYSU-A dataset are summarized in Table~\ref{tab:UVFP}, with the corresponding confusion matrix depicted in Figure~\ref{fig:cm_uvfp}. With the integration of all enhanced modules, our system achieves impressive performance, with all metrics surpassing the 82\% threshold, affirming its proficiency in distinguishing left from right VFP. An ablation study further corroborates these findings. The table illustrates that the adoption of QF techniques significantly hones classification accuracy, as reflected by the enhancement in performance metrics. The addition of diffusion-based refinement also contributes to superior classification outcomes, excelling beyond the baseline model across all performance metrics. Moreover, the confusion matrix (Figure~\ref{fig:cm_uvfp}) clearly shows the benefits of integrating the two enhanced modules. In comparison to the baseline model, which demonstrates a high rate of misclassification from left to right VFP and a lower rate in the opposite direction, the system predominately predicts right VFP over left VFP. However, this bias is substantially mitigated by the integration of both QF and diffusion-based refinement techniques, showing a better classification result.

To further illustrate the efficacy of our system, we replace the oracle VFP labels with predictions from the initial multimodal VFP model~(outlined in Section~\ref{sec:paralysis_analysis} and results shown in Section~\ref{subsec:vfp_performance}). Subsequently, in the second phase, we differentiate the system VFP cases into left or right VFP by analyzing the variance between LVFDyn and RVFDyn. The confusion matrices depicted in Figure~\ref{fig:cm_uvfp_pipe} present the comprehensive three-category classification outcomes. The development of the multimodal VFP model, we randomly make train-test splits, yielding 445 cases~(81 normal, 364 VFP) for training the detection model, and 75 cases~(25 normal, 25 left VFP, 25 right VFP) for testing. As depicted in Figure~\ref{fig:cm_uvfp_pipe}, the majority of the samples were correctly identified, with the largest numbers in the confusion matrices forming their diagonals, showing an accuracy of 81.3\%.

\subsection{Auxiliary Visual Analysis for UVFP}
In this subsection, the potential of the novel feature is demonstrated through expert visual analysis. Figure~\ref{fig:analysis_results} showcases the GAW and VFDyn plots for a patient diagnosed with right VFP from the SYSU-A dataset. The MLVAS system extracts several video highlights, yielding multiple analytical outcomes. The left three figures present the time-normalized GAW, illustrating the global vibratory behavior of the vocal folds during phonation. However, these plots are not capable of differentiating between the movements of the individual vocal folds, thus clinicians are unable to ascertain the paralyzed vocal fold based solely on the GAW. In contrast, the VFDyn analysis, depicted in the figures listed vertically on the right, offers a disambiguated view of each vocal fold's motion. This granular perspective significantly augments the utility of the metrics provided by the MLVAS for clinical assessment of UVFP. In this specific example, the VFDyn plots display a markedly greater level of oscillation for the LVFDyn in comparison to the RVFDyn, which is consistent with the patient's right VFP diagnosis, thus validating the clinical utility of the MLVAS-derived metrics.

\section{Conclusion}
\label{sec:conclusion}
In this work, we present the Multimodal Laryngoscopic Video Analyzing System~(MLVAS), a novel framework that integrates audio and image processing for the advanced diagnosis of laryngeal disorders. The front-end utilizes an audio-based KWS model to capture the full phonation cycle, along with a vocal fold detection module to confirm the presence of vocal folds. Leveraging HSV color space analysis, MLVAS extracts stroboscopic sequences, enabling the automatic generation of video highlights from raw laryngoscopic footage, thereby eliminating the need for manual pre-editing. The back-end integrates both audio and visual modalities for VFP detection, marking the first application of pre-trained audio models for laryngeal disease diagnosis to encode robust audio features. Visual metrics are computed using a two-stage segmentation pipeline, combining U-Net-based segmentation with diffusion refinement to reduce false positives, followed by quadratic fitting to derive LVFDyn and RVFDyn, representing the dynamic angle deviations of the left and right vocal folds. The novel visual metrics enable MLVAS not only to detect VFP but also to distinguish left- from right-sided VFP. By uniting audio and visual features in a cohesive framework, MLVAS achieves superior performance on the clinical dataset and provides interpretable visualizations, such as the GAW and VFDyn, equipping clinicians with reliable tools for precise and efficient UVFP diagnosis.

\section{Acknowledgment}
This research is funded by the DKU foundation project ``Interdisciplinary Signal Processing Technologies''. Many thanks for the computational resource provided by the Advanced Computing East China Sub-Center.

\bibliographystyle{elsarticle-num} 
\bibliography{ref}

\end{document}